\documentclass{ws-procs975x65}

\def\ltap{\ \raisebox{-.4ex}{\rlap{$\sim$}} \raisebox{.4ex}{$<$}\ }
\def\gtap{\ \raisebox{-.4ex}{\rlap{$\sim$}} \raisebox{.4ex}{$>$}\ }

\newcommand{\betabeta}{\mbox{$(\beta \beta)_{0 \nu}  $}}

\newcommand{\meff}{\mbox{$\left|  < \!  m \!  > \right| \ $}}

\newcommand{\hbeta}{$\mbox{}^3 {\rm H}$ $\beta$-decay \ }
\newcommand{\eV}{\mbox{$ \  \mathrm{eV} \ $}}

\newcommand{\pmns}{\mbox{$ U_{\rm PMNS}$}}
\newcommand{\bea}{\begin{equation}\begin{array}{c}}
\newcommand{\eea}{\end{array}\end{equation}}
\newcommand{\ea}{\end{array}} 

\newcommand{\beq}{\begin{equation}}
\newcommand{\eeq}{\end{equation}}
\newcommand{\bad}{\begin{array}{ccc}}

\newcommand{\mefff}{\mbox{$ < \! m \! > $}}
\newcommand{\ba}{\begin{array}{c}}
\let\trimmarks\relax
\begin{document}

\vspace*{-15mm}
\begin{flushright}
Ref. SISSA 22/2014/FISI
\end{flushright}
\vspace*{0.7cm}

\title{LEPTONIC CP VIOLATION AND LEPTOGENESIS}

\author{S. T. PETCOV 
\footnote{Invited talk given at 
the Conference in Honour of 
the 90th Birthday of Freeman Dyson, Nanyang Technological University, 
Singapore, 26-29 August 2013; 
published in the Proceedings of the Conference 
(Eds. K K Phua {\it et al.},
World Scientific, 2014), p. 179,
and in
Int. J. Mod. Phys. A  {\bf 29} (2014) 1430028.}
}

\address{SISSA/INFN, Trieste, Italy, and\\
Kavli IPMU, University of Tokyo, Tokyo, Japan} 
%
%
%
%
%

\begin{abstract}
The phenomenology of 3-neutrino mixing, 
the current status of our knowledge about the 
3-neutrino mixing parameters, including 
the absolute neutrino mass scale, 
and of the Dirac and Majorana CP violation 
in the lepton sector, are reviewed. 
The problems of CP violation in neutrino 
oscillations and of determining the nature -
Dirac or Majorana - of massive neutrinos, 
are discussed. The seesaw mechanism of neutrino 
mass generation and the related leptogenesis 
scenario of generation of the baryon asymmetry 
of the Universe, are considered.
The results showing that the CP violation 
necessary for the generation 
of the baryon asymmetry of the Universe
in leptogenesis can be due exclusively to the Dirac 
and/or Majorana CP-violating phase(s) in the 
neutrino mixing matrix $U$, are briefly reviewed. 
\end{abstract}


\vspace{-0.1cm}
\bodymatter
%
\section{Introduction: Neutrinos (Preliminary Remarks)}
\label{aba:sec1}
%

 It is both an honor and a pleasure 
to speak at this Conference,
organized in honor of Prof. Freeman Dyson's 
90th birthday. 
%
 My talk will be devoted to aspects of neutrino physics, 
so I would like to start by recalling some basic 
facts about neutrinos 
\footnote{In this introductory part of the article 
I am following the reviews\cite{PDG2012,STPNuNature2013}~.
}. 
It is well established experimentally 
that the neutrinos and antineutrinos
which take part in the standard
charged current (CC) and neutral
current (NC)  weak interaction
are of three varieties (types) or flavours:
electron, $\nu_e$ and $\bar{\nu}_e$,
muon, $\nu_\mu$ and $\bar{\nu}_\mu$, and
tauon, $\nu_\tau$ and  $\bar{\nu}_\tau$.
The notion of neutrino type or flavour
is dynamical: $\nu_e$ is the neutrino
which is produced with $e^+$, or produces an $e^-$
in CC weak interaction processes;
$\nu_\mu$ is the neutrino which
is produced with $\mu^+$, or
produces  $\mu^-$, etc.
The flavour of a given neutrino
is Lorentz invariant.
Among the three different flavour neutrinos and
antineutrinos, no two are identical.
Correspondingly, the states which describe
different flavour neutrinos must be orthogonal
(within the precision of the current data):
$\langle \nu_{l'} |\nu_l\rangle = \delta_{l'l}$,
$\langle \bar{\nu}_{l'} |\bar{\nu}_l\rangle = \delta_{l'l}$,
$\langle \bar{\nu}_{l'} |\nu_l\rangle = 0$.

  It is also well-known from the existing data
(all neutrino experiments were done
so far with relativistic neutrinos
or antineutrinos), that the flavour
neutrinos $\nu_l$ (antineutrinos $\bar{\nu}_l$), are
always produced in weak interaction
processes in a state that is
predominantly left-handed (LH)
(right-handed (RH)).
To account for this fact,
$\nu_l$ and $\bar{\nu}_l$
are described in the
Standard Theory (ST) by a chiral LH flavour
neutrino field $\nu_{lL}(x)$, $l=e,\mu,\tau$.
For massless $\nu_l$, the
state of $\nu_l$ ($\bar{\nu}_l$) which
the  field  $\nu_{lL}(x)$ annihilates (creates)
is with helicity (-1/2) (helicity +1/2).
If $\nu_l$ has a non-zero mass $m(\nu_l)$,
the state of $\nu_l$ ($\bar{\nu}_l$)  is a linear
superposition of the helicity (-1/2) and (+1/2)
states, but the helicity +1/2 state (helicity (-1/2) state)
enters into the superposition with a coefficient $\propto m(\nu_l)/E$,
E being the neutrino energy, and thus is strongly
suppressed. Together with the LH
charged lepton field $l_{L}(x)$, $\nu_{lL}(x)$ forms
an $SU(2)_{L}$ doublet. In the absence of neutrino
mixing and zero neutrino masses,
$\nu_{lL}(x)$ and $l_{L}(x)$ can be assigned
one unit of the additive lepton charge
$L_l$, and the three charges $L_l$, $l=e,\mu,\tau$,
as well as the total lepton charge, 
$L = L_e + L_{\mu} + L_{\tau}$,
are conserved by the weak interaction.

  At present there is no compelling evidence
for the existence of states of relativistic neutrinos
(antineutrinos), which are predominantly
right-handed, $\nu_R$ (left-handed, $\bar{\nu}_L$).
If RH neutrinos and LH antineutrinos exist, their interaction with matter
should be much weaker than the weak interaction of the flavour LH neutrinos
$\nu_l$ and RH antineutrinos $\bar{\nu}_l$, i.e., $\nu_R$ ($\bar{\nu}_L$)
should be ``sterile'' or ``inert'' neutrinos
(antineutrinos)\cite{BPont67}. In the formalism of the Standard
Theory, the sterile $\nu_R$ and $\bar{\nu}_L$
can be described by $SU(2)_L$ singlet RH neutrino 
fields $\nu_R(x)$. In this case, $\nu_R$ and $\bar{\nu}_L$
will have no gauge interactions, i.e., 
will not couple to the weak $W^{\pm}$
and $Z^0$ bosons. The simplest hypothesis 
(based on symmetry considerations)
is that to each LH flavour
neutrino field $\nu_{lL}(x)$ there
corresponds a RH neutrino
field $\nu_{lR}(x)$, $l=e,\mu,\tau$, 
although schemes with less (more) than three 
RH neutrinos are also being considered.

If present in an
extension of the Standard Theory 
(even in the minimal one),
the RH neutrinos can play a crucial role
i) in the generation
of neutrino masses and mixing,
ii) in understanding
the remarkable disparity between
the magnitudes of neutrino masses and
the masses of the charged leptons and quarks,
and iii) in the generation of the observed
matter-antimatter asymmetry of the
Universe (via the leptogenesis
mechanism\cite{LGFY,kuzmin}).  In this scenario
which is based on the see-saw theory\cite{seesaw},
there is a link between the generation
of neutrino masses and the generation of the
matter-antimatter (or baryon) asymmetry  of the Universe.
In this talk we will review 
this remarkable connection.
We will discuss also the interesting possibility 
that the CP violation necessary for the generation of the 
observed matter-antimatter asymmetry of the Universe
in the leptogenesis scenarion of the asymmetry generation 
can be provided exclusively by 
the Dirac and/or Majorana~\cite{BHP80}~ CP violation phases,  
present in the neutrino mixing matrix~\cite{PPRio106,EMSTP09}~. 

\vspace{-0.4cm}
%
\section{The Neutrino Mixing}
\label{aba:sec2}
%

There have been remarkable discoveries in the field of neutrino 
physics in the last 15 years or so.
The experiments with solar, atmospheric, 
reactor and accelerator neutrinos
have provided compelling evidences for the
existence of neutrino oscillations\cite{BPont57,MNS62}~, 
transitions in flight between the different flavour neutrinos
$\nu_e$, $\nu_\mu$, $\nu_\tau$
(antineutrinos $\bar{\nu}_e$, $\bar{\nu}_\mu$, $\bar{\nu}_\tau$),
caused by nonzero neutrino masses and neutrino mixing 
(see, e.g., Ref.~\refcite{PDG2012} for review of the relevant data).
The existence of flavour neutrino oscillations
implies that if a neutrino of a given flavour, say
$\nu_\mu$, with energy $E$ is produced in some
weak interaction process, at a sufficiently
large distance $L$ from the $\nu_\mu$ source
the probability to find a neutrino of a different flavour,
say $\nu_\tau$, $P(\nu_\mu\rightarrow \nu_\tau;E,L)$,
is different from zero.
$P(\nu_\mu\rightarrow \nu_\tau;E,L)$ is
called the $\nu_\mu\rightarrow \nu_\tau$
oscillation or transition probability.
If $P(\nu_\mu\rightarrow \nu_\tau;E,L) \neq 0$,
the probability that $\nu_\mu$ will not 
change into a neutrino of a different flavour,
i.e., the ``$\nu_\mu$ survival probability''
$P(\nu_\mu\rightarrow \nu_\mu;E,L)$,
will be smaller than one. If only muon neutrinos $\nu_\mu$
are detected in a given experiment and they take 
part in oscillations, one would observe a ``disappearance''
of muon neutrinos on the way from the
$\nu_\mu$ source to the detector.

The existing data, accumulated over 
more than 15 years allowed to firmly establish the  
existence of oscillations of the solar $\nu_e$ 
($E \cong (0.23 - 14.4)$ MeV), 
atmospheric $\nu_{\mu}$ and $\bar{\nu}_{\mu}$ 
($E\cong (0.2 - 100)$ GeV) crossing the Earth, 
accelerator $\nu_{\mu}$ ($E\sim 1$ GeV)
at $L= 250;~295;~730$ km 
and reactor $\bar{\nu}_e$ ($E \cong (2.6-10.0)$ MeV) 
at $L\sim 1;~180$ km. The data imply the presence of 
mixing in the weak charged  lepton current:
\begin{equation}
\label{CC}
{\cal L}_{\rm CC} = - ~\frac{g}{\sqrt{2}}\,
\sum_{l=e,\mu,\tau}
\overline{l_L}(x)\, \gamma_{\alpha} \nu_{lL}(x)\,
W^{\alpha \dagger}(x) + h.c.\,,~
\nu_{l \mathrm{L}}(x)
= \sum^n_{j=1} U_{l j} \nu_{j \mathrm{L}}(x)\,,
\end{equation}
%
\noindent where 
$\nu_{lL}(x)$ are the flavour neutrino fields, 
$\nu_{j \mathrm{L}}(x)$ is the left-handed (LH)
component of the field of the neutrino $\nu_j$ having a 
mass $m_j$, and $U$ is a unitary matrix - the
Pontecorvo-Maki-Nakagawa-Sakata (PMNS)
neutrino mixing matrix\cite{BPont57,MNS62,BPont67}~, 
$U\equiv U_{PMNS}$.
All compelling  neutrino oscillation data
can be described assuming 3-neutrino mixing in vacuum, $n=3$.
The number of massive neutrinos $n$ 
can, in general, be bigger than 3 if, e.g., 
there exist RH
sterile neutrinos\cite{BPont67}
and they mix with the LH flavour neutrinos. 
It follows from the current data that at least 3 of 
the neutrinos $\nu_j$, say 
$\nu_1$, $\nu_2$, $\nu_3$, must be light,
$m_{1,2,3} \ltap 1$ eV, and must have different 
masses, $m_1\neq m_2 \neq m_3$  
\footnote{At present there are several 
experimental inconclusive hints 
for existence of one or two light 
sterile neutrinos at the eV scale, 
which mix with the flavour neutrinos, 
implying the presence in the neutrino mixing  
of additional one or two neutrinos, $\nu_4$ or $\nu_{4,5}$, 
with masses $m_4~(m_{4,5})\sim 1~{\rm eV}$ 
(see, e.g.,  Refs.~\refcite{SterNuWhitePaper,SnowM2013}).   
The discussion of these hints and of the 
related implications is out of the scope of the 
present article.}.

 In the case of 3 light neutrinos, the 
neutrino mixing matrix $U$ can be 
parametrised by 3 angles and, 
depending on whether 
the  massive neutrinos $\nu_j$ are Dirac 
or Majorana particles, 
by one Dirac, or one Dirac and two Majorana,
CP violation (CPV) phases\cite{BHP80}~: 
\begin{equation}
U= VP\,,~~~
P = {\rm diag}(1, e^{i \frac{\alpha_{21}}{2}}, e^{i \frac{\alpha_{31}}{2}})\,, 
\label{VP}
\end{equation}
%
where $\alpha_{21,31}$ 
are two Majorana CPV 
phases and $V$ is a CKM-like matrix, 
\begin{equation} 
\begin{array}{c}
\label{eq:Vpara}
V = \left(\begin{array}{ccc} 
 c_{12} c_{13} & s_{12} c_{13} & s_{13} e^{-i \delta}  \\[0.2cm] 
 -s_{12} c_{23} - c_{12} s_{23} s_{13} e^{i \delta} 
 & c_{12} c_{23} - s_{12} s_{23} s_{13} e^{i \delta} 
 & s_{23} c_{13} 
\\[0.2cm] 
 s_{12} s_{23} - c_{12} c_{23} s_{13} e^{i \delta} & 
 - c_{12} s_{23} - s_{12} c_{23} s_{13} e^{i \delta} 
 & c_{23} c_{13} 
\\ 
  \end{array}    
\right)\,. 
\end{array} 
\end{equation}
%
\noindent 
In Eq.~(\ref{eq:Vpara}),
$c_{ij} = \cos\theta_{ij}$, 
$s_{ij} = \sin\theta_{ij}$,
the angles $\theta_{ij} = [0,\pi/2]$, 
and $\delta = [0,2\pi)$ is the 
Dirac CPV phase. 
Thus, in the case of massive Dirac neutrinos, 
the neutrino mixing matrix $U$ is similar,
in what concerns the number of 
mixing angles and CPV phases, to the CKM quark 
mixing matrix. The presence of two additional 
physical CPV phases in $U$ if $\nu_j$ 
are Majorana particles is a consequence of 
the special properties  of the latter
(see, e.g., Refs.~\refcite{STPNuNature2013,BiPet87,BHP80}).
On the basis of the existing neutrino data 
it is impossible to determine whether the massive 
neutrinos are Dirac or Majorana fermions.

 The  neutrino oscillation probabilities depend
on the neutrino energy, $E$, 
the source-detector distance $L$,
on the elements of $U$ and, 
for relativistic neutrinos used in all neutrino 
experiments performed so far, on the neutrino mass squared 
differences $\Delta m^2_{ij} \equiv (m^2_{i} - m^2_j)$, $i\neq j$ 
(see, e.g., Ref.~\refcite{BiPet87}).
In the case of 3-neutrino mixing
there are only two independent $\Delta m^2_{ij}$,
say $\Delta m^2_{21}\neq 0$ and $\Delta m^2_{31} \neq 0$.
The numbering of the neutrinos $\nu_j$ is arbitrary.
We will employ  the widely used convention 
which allows to associate $\theta_{13}$ with the smallest 
mixing angle in the PMNS matrix, 
and  $\theta_{12}$, $\Delta m^2_{21}> 0$, and
$\theta_{23}$, $\Delta m^2_{31}$,
with the parameters which drive the 
solar ($\nu_e$) and the dominant atmospheric 
$\nu_{\mu}$ and $\bar{\nu}_{\mu}$ 
oscillations, respectively.
In this convention $m_1 < m_2$,
$ 0 < \Delta m^2_{21} < |\Delta m^2_{31}|$, 
and, depending on ${\rm sgn}(\Delta m^2_{31})$,
we have either $m_3 < m_1$ or $m_3 > m_2$. 
The existing  data allow us to determine 
$\Delta m^2_{21}$, $\theta_{12}$, and
$|\Delta m^2_{31(32)}|$, $\theta_{23}$ and $\theta_{13}$, 
with a relatively good 
precision\cite{Capozzi:2013csa,CGGMSchw12update}~. 
The best fit values and the 3$\sigma$ allowed ranges of 
$\Delta m^2_{21}$, $s^2_{12}$, $|\Delta m^2_{31(32)}|$, 
$s^2_{23}$ and $s^2_{13}$ read\cite{Capozzi:2013csa}~:
\vspace{-0.1cm}
\begin{eqnarray}
\label{deltasolvalues}
&(\Delta m^2_{21})_{\rm BF} = 7.54 \times 10^{-5}\ \eV^2,~~~~~~ 
 \Delta m^2_{21} = (6.99 - 8.18) \times 10^{-5} \ \eV^2\,,\\
\label{sinsolvalues}
& (\sin^2 \theta_{12})_{\rm BF} = 0.308,~~~~~~
 0.259 \leq \sin^2 \theta_{12} \leq 0.359\,,
~~~~~~~~~~~~~~~~~~~~~~~~\\ 
\label{deltaatmvalues}
&(|\Delta m^2_{31(32)}|)_{\rm BF} = 2.48~(2.44) \times 10^{-3} \ \eV^2\,,
~~~~~~~~~~~~~~~~~~~~~~~~~~~~~~~~~~~~~~~~~~~\\ 
& |\Delta m^2_{31(32)}| = (2.26~(2.21) - 2.70~(2.65))\times 10^{-3}\ \eV^2\,,
~~~~~~~~~~~~~~~~~~~~~~~~~~~~~~~~~~~~~\\ 
\label{thetaatmvalues}
&(\sin^2\theta_{23})_{\rm BF} = 0.425~(0.437)\,,~~ 
 0.357(0.363) \leq \sin^2\theta_{23} \leq 0.641(0.659)\,,
~~~~~~~~~\\
\label{theta13values}
&(\sin^2\theta_{13})_{\rm BF} = 0.0234~(0.0239)\,,~~ 
0.0177(0.0178) \leq \sin^2\theta_{23} \leq 0.0297(0.300)\,,
~~~~~~~~~
\end{eqnarray}
%
where when there are two values one of which is in brackets, 
the value (the value in brackets) 
corresponds to  $\Delta m^2_{31(32)}>0$ ($\Delta m^2_{31(32)} <0$).
There are also hints from data about the value of 
the Dirac phase $\delta$. In both 
analyses\cite{Capozzi:2013csa,CGGMSchw12update}
the authors find that the best fit value of 
$\delta \cong 3\pi/2$. 
The CP conserving values $\delta = 0$ and $\pi$ 
($\delta = 0)$ are 
disfavored at $1.6\sigma$ to  $2.0\sigma$ 
(at $2.0\sigma$) for  $\Delta m^2_{31(32)}>0$
($\Delta m^2_{31(32)}<0$). 
In the case of $\Delta m^2_{31(32)}<0$, the value 
$\delta = \pi$ is statistically $1\sigma$ away from 
the best fit value $\delta \cong 3\pi/2$ 
(see, e.g., Fig. 3 in Ref.~\refcite{Capozzi:2013csa}).

Thus, we have $\Delta m^2_{21}/|\Delta m^2_{31(32)}| \cong 0.03$, and
$|\Delta m^2_{31}| = |\Delta m^2_{32} - \Delta m^2_{21}| 
\cong |\Delta m^2_{32}|$.
Maximal solar neutrino mixing, i.e. 
$\theta_{12} = \pi/4$, is ruled out at more than 
6$\sigma$ by the data. Correspondingly, one has  
$\cos2\theta_{12} \geq 0.28$ (at $99.73\%$ C.L.).
The angle $\theta_{13}$ was measured relatively recently - 
in the spring of 2012 - in the high precision 
Daya Bay\cite{DBayth13} and RENO\cite{RENOth13} experiments.

The results quoted above imply also that
$\theta_{23} \cong \pi/4$, $\theta_{12} \cong \pi/5.4$ and 
that $\theta_{13} < \pi/13$. Correspondingly, the pattern of 
neutrino mixing is drastically different from 
the pattern of quark mixing. 

The existing data do not allow one to 
determine the sign of $\Delta m^2_{31(32)}$.
In the case of 3-neutrino mixing,  
the two possible signs of
$\Delta m^2_{31(32)}$ correspond to two 
types of neutrino mass spectrum.
In the convention of numbering 
the neutrinos $\nu_j$ 
employed by us, the two spectra read:\\
{\it i) spectrum with normal ordering (NO)}:
$m_1 < m_2 < m_3$, $\Delta m^2_{31(32)} >0$,
$\Delta m^2_{21} > 0$,
$m_{2(3)} = (m_1^2 + \Delta m^2_{21(31)})^{1\over{2}}$; \\~~
{\it ii) spectrum with inverted ordering (IO)}:
$m_3 < m_1 < m_2$, $\Delta m^2_{32(31)}< 0$, 
$\Delta m^2_{21} > 0$,
$m_{2} = (m_3^2 + \Delta m^2_{23})^{1\over{2}}$, 
$m_{1} = (m_3^2 + \Delta m^2_{23} - \Delta m^2_{21})^{1\over{2}}$.\\ 
Depending on the values of the lightest neutrino mass, 
${\rm min}(m_j)$, the neutrino mass spectrum can also be:~\\
{\it a) Normal Hierarchical (NH)}: $m_1 \ll m_2 < m_3$, 
$m_2 \cong (\Delta m^2_{21})^
{1\over{2}}\cong 8.7\times 10^{-3}$ eV,
$m_3 \cong (\Delta m^2_{31})^{1\over{2}} \cong 0.050$ eV; or\\ 
{\it b) Inverted Hierarchical (IH)}: $m_3 \ll m_1 < m_2$, 
$m_{1,2} \cong |\Delta m^2_{32}|^{1\over{2}}\cong 0.049$ eV; or\\ 
{\it c) Quasi-Degenerate (QD)}: $m_1 \cong m_2 \cong m_3 \cong m_0$,
$m_j^2 \gg |\Delta m^2_{31(32)}|$, $m_0 \gtap 0.10$ eV.\\ 
All three types of spectrum are compatible
with the existing constraints on the absolute scale 
of neutrino masses $m_j$. Determining the type of neutrino 
mass spectrum is one of the main goals of the future 
experiments in the field of neutrino physics
\footnote{For a brief discussion of experiments 
which can provide data on the type of 
neutrino mass spectrum see, e.g., Ref.~\refcite{PPNH07};
for some specific proposals see, e.g., Ref.~\refcite{NOIO}.}
(see, e.g., Refs.~\refcite{PDG2012,SnowM2013,Cahn2013}).

  Information about the 
absolute neutrino mass scale 
(or about ${\rm min}(m_j)$)
can be obtained, e.g., by measuring the spectrum 
of electrons near the end point in 
\hbeta experiments\cite{Fermi34,Mainz,MoscowH3}
and from cosmological and astrophysical data. 
The most stringent upper 
bounds on the $\bar{\nu}_e$ mass 
were obtained in the Troitzk~\cite{MoscowH3b} 
experiment:
\beq
m_{\bar{\nu}_e} < 2.05~\mathrm{eV}~~~\mbox{at}~95\%~\mathrm{C.L.} 
\label{H3beta}
\eeq
%
\noindent Similar result was obtained in the 
Mainz experiment\cite{Mainz}~: 
$m_{\bar{\nu}_e} < 2.3~\rm{eV}$ at 95\% CL.
We have $m_{\bar{\nu}_e} \cong m_{1,2,3}$
in the case of QD  spectrum. 
The KATRIN experiment~\cite{MainzKATRIN}
is planned to reach sensitivity  
of  $m_{\bar{\nu}_e} \sim 0.20$~eV,
i.e., it will probe the region of the QD 
spectrum. 

  The Cosmic Microwave Background (CMB)
data of the WMAP experiment, combined with
supernovae data and data on galaxy clustering
can be used to obtain an upper limit on the sum of
neutrinos masses (see and, e.g., Ref.~\refcite{summj}).
Depending on the model complexity and the input data used 
one obtains\cite{summj}~:
$\sum_j m_j\ltap (0.3 - 1.3)$ eV, 95\% CL.

 In March of 2013 the Planck Collaboration 
published their first constraints on 
$\sum_j m_j$\cite{Ade:2013lta}~. 
Assuming the existence of three 
massive neutrinos and the validity of the $\Lambda$ CDM 
(Cold Dark Matter) model, and combining their data 
on the CMB temperature power spectrum 
with the WMAP polarisation low-multiple 
($\ell\leq$ 23) and ACT high-multiple ($\ell\geq$ 2500)
CMB data\cite{WMAPascitedbyPlanck,ACTascitedbyPlanck}~,
the Planck Collaboration reported the following 
upper limit on the sum of the neutrino 
masses\cite{Ade:2013lta}~:
\begin{equation}
\sum_j m_j\, < \, 0.66~eV,~~~ 95\%~{\rm CL.}
\label{Planck1}
\end{equation}
%
Adding the data on the Baryon Acoustic 
Oscillations (BAO) lowers significantly the 
limit\cite{Ade:2013lta}~:
$\sum_j m_j\, < \, (0.23$ eV), 95\% CL.

It follows from these data that neutrino masses are
 much smaller than the masses of
 charged leptons and quarks.
 If we take as an indicative
 upper limit $m_j \ltap 0.5$ eV, we have
 $m_j/m_{l,q} \ltap 10^{-6}$, where 
$m_l$ and $m_q$ are the charged lepton 
and quark masses, $l=e,\mu,\tau$,
 $q=d,s,b,u,c,t$. It is natural to suppose that
 the remarkable smallness of neutrino masses
 is related to the existence of a new
 fundamental mass scale in particle physics,
 and thus to new physics beyond that predicted by the
 Standard Theory.

\vspace{-0.2cm}

%
\section{CP Violation in the Lepton Sector}
%

%
\subsection{Dirac CP Violation}
%

 The relatively large value of 
$\sin\theta_{13} \cong 0.15$ 
measured with a high precision in the Daya Bay\cite{DBayth13}
and RENO\cite{RENOth13} experiments has  
far-reaching implications for the program of 
research in neutrino physics, and more specifically,\\
i) for the determination of the type of neutrino mass spectrum (or of
${\rm sgn}(\Delta m^2_{31(32)}$)) in neutrino oscillation
experiments (see, e.g., Refs. \refcite{NOIO,Cahn2013});\\
ii) for understanding the pattern of the neutrino mixing
and its origins (see, e.g., Ref.~\refcite{Marzocca:2013cr}
and the references quoted therein);\\
iii) for the predictions for the $\betabeta$-decay 
effective Majorana mass in the case of NH light 
neutrino mass spectrum (see, e.g., Ref.~\refcite{PPNH07}).

 The relatively large value of $\sin\theta_{13} \cong 0.15$  
combined with the value of $\delta = 3\pi/2$ 
has far-reaching implications for the searches for 
CP violation in neutrino oscillations (see further).
It has also important implications for the ``flavoured'' 
leptogenesis scenario of generation of the 
baryon asymmetry of the Universe (BAU).
As we will discuss in greater detail in Section 5, 
if all CP violation necessary for the generation of BAU 
is due to the Dirac phase $\delta$, 
a necessary condition for reproducing 
the observed BAU is\cite{PPRio106} 
$|\sin\theta_{13}\,\sin\delta|\gtap 0.09$, which is comfortably 
compatible with the measured value of $\sin\theta_{13}$ and with 
best fit value of $\delta \cong 3\pi/2$.  

 A CP nonconserving value of the Dirac phase $\delta$ 
will cause CP violation in flavour neutrino oscillations, 
$\nu_l \rightarrow \nu_{l'}$, $\bar{\nu}_l \rightarrow \bar{\nu}_{l'}$, 
$l\neq l'=e,\mu,\tau$. Indeed,  CP-, T- and CPT- invariance 
imply for $\nu_l \rightarrow \nu_{l'}$ 
oscillation probabilities\cite{Cabibbo78,BHP80}~:
\begin{eqnarray}
\label{CP}
P(\nu_l \rightarrow\nu_{l'}) =
P(\bar{\nu}_{l}{\small\rightarrow}\bar{\nu}_{l'})\,,
~~{\rm CP-invariance}\,,\\[0.30cm]
\label{Tnu}
P(\nu_l{\small\rightarrow}\nu_{l'})=
P(\nu_{l'}{\small\rightarrow}\nu_{l})\,,~~{\rm T-invariance}\,,\\[0.30cm]
\label{Tantinu}
P(\bar{\nu}_l{\small\rightarrow}\bar{\nu}_{l'})=
P(\bar{\nu}_{l'}{\small\rightarrow}\bar{\nu}_{l})\,,
~~{\rm T-invariance}\,,\\[0.30cm]
P(\nu_l \rightarrow \nu_{l'})=
P(\bar{\nu}_{l'} \rightarrow \bar{\nu}_{l})\,,~~{\rm CPT-invariance}\,,
\end{eqnarray}
%
where  $l,l'=e,\mu,\tau$. 
It follows from CPT-invariance that for  $l=l'=e,\mu,\tau$ 
we have: 
\begin{equation}
P(\nu_l \rightarrow \nu_{l})= P(\bar{\nu}_{l} \rightarrow \bar{\nu}_{l})\,.
\label{CPTsurv}
\end{equation}
%
From the comparison of Eqs. (\ref{CP}) and (\ref{CPTsurv}) 
it is clear that if CPT invariance holds, which we will 
assume to be the case, the ``disappearance'' neutrino oscillation 
experiments in which one gets information about the probabilities 
$P(\nu_l \rightarrow \nu_{l})$ and 
$P(\bar{\nu}_{l} \rightarrow \bar{\nu}_{l})$, $l=e,\mu,\tau$,
are not sensitive to CP-violation.
Therefore, a measure of $CP$- and $T$- violation is provided
by the asymmetries\cite{BHP80,PKSP3nu88,Barger:1980jm}~:
\begin{eqnarray}
A^{(l,l')}_{\rm CP} = P({\nu_l \rightarrow \nu_{l'}}) 
- P({\bar{\nu}_l \rightarrow \bar{\nu}_{l'}})\,,~~
l\neq l'=e,\mu,\tau\,,\\[0.30cm] 
A^{(l,l')}_{\rm T} = P(\nu_l \rightarrow \nu_{l'}) -
P(\nu_{l'} \rightarrow \nu_{l})\,,~~l\neq l'=e,\mu,\tau\,.
\end{eqnarray}
%
\noindent For 3-$\nu$ oscillations in vacuum one has\cite{PKSP3nu88}~:\\
\begin{eqnarray}
\label{CPTAsym}
A^{(e,\mu)}_{\rm CP} = A^{(\mu,\tau)}_{\rm CP} = 
- A^{(e,\tau)}_{\rm CP} = A^{(e,\mu)}_{\rm T} = A^{(\mu,\tau)}_{\rm T} = - A^{(e,\tau)}_{\rm T} = 
 J_{\rm CP}~F^{vac}_{osc}\,,\\[0.30cm]
\label{JCP}
J_{CP} = {\rm Im}\left\{ U_{e1} U_{\mu 2} U_{e 2}^\ast 
U_{\mu 1}^\ast \right\} = \frac{1}{8}\sin2\theta_{12} \sin2\theta_{23}
\sin2\theta_{13}\cos\theta_{13}\sin\delta\,,\\[0.30cm]
\label{CPVoscF}
F^{vac}_{osc}=
\sin(\frac{\Delta m^2_{21}}{2E}L) +
\sin(\frac{\Delta m^2_{32}}{2E}L) +
\sin(\frac{\Delta m^2_{13}}{2E}L)\,.
\end{eqnarray}
%
Thus, the magnitude of  CP violation 
effects in neutrino oscillations 
is controlled  by the rephasing invariant  
associated with the Dirac phase $\delta$, 
$J_{\rm CP}$~\cite{PKSP3nu88}~. 
The latter is analogous to the rephasing invariant  
associated with the 
Dirac phase in the Cabibbo-Kobayashi-Maskawa
quark mixing matrix, introduced in Ref.~\refcite{CJ85}~.
The existence of Dirac CPV in the lepton sector 
would be established if, e.g., some 
of the vacuum oscillation asymmetries 
$A^{(e,\mu)}_{\rm CP(T)}$, $A^{(e,\tau)}_{\rm CP}$, etc. 
are proven experimentally to be nonzero.
This would imply that $J_{\rm CP} \neq 0$,
and, consequently, that $\sin\theta_{13}\sin\delta \neq 0$, 
which in turn would mean that 
$\sin\delta \neq 0$ since $\sin\theta_{13}\neq 0$.

 Given the fact that $\sin 2\theta_{12}$,
$\sin 2\theta_{23}$ and $\sin 2\theta_{13}$ have been 
determined experimentally with a relatively 
good precision, the size of CP violation
effects in neutrino oscillations depends
essentially only on the magnitude 
of the currently not well determined 
value of  the Dirac phase $\delta$.
The current data implies $|J_{CP}| \ltap  0.038\,|\sin\delta|$,
where we have used the $3\sigma$ ranges of 
$\sin^2\theta_{12}$, $\sin^2\theta_{23}$ and $\sin^2\theta_{13}$
given in Eqs. (\ref{deltasolvalues}) - (\ref{theta13values}).
For the best fit values of 
$\sin^2\theta_{12}$, $\sin^2\theta_{23}$ and $\sin^2\theta_{13}$
and $\delta$ we find in the case of $\Delta m^2_{31(32)} > 0$ 
($\Delta m^2_{31(32)} < 0$):  $J_{CP} \cong -\,0.032\,(-\,0.031)$.
Thus, if the indication that $\delta \cong 3\pi/2$ 
is confirmed by future more precise data, 
the CP violation effects in neutrino 
oscillations would be relatively large if the factor 
$F^{vac}_{osc}$ is not suppressing the CPV asymmetries.
We would have  $F^{vac}_{osc} \cong 0$
and the CPV asymmetries will be strongly suppressed,
as it follows from Eqs.~(\ref{CPTAsym}) and (\ref{CPVoscF}),
if under the conditions of a given experiment 
one of the two neutrino mass squared differences, 
say $\Delta m^2_{21}$, is not operative, i.e., 
$\sin(\Delta m^2_{21}L/(2E))\cong 0$.
In this case the CP violation effects in neutrino oscillations 
will be hardly observable. 

One of the major goals of the future 
experimental studies in neutrino physics
is the searches for CPV effects due to the Dirac 
phase in the PMNS mixing matrix
(see, e.g., Refs.~\refcite{SnowM2013,LBLFuture13}).
It follows from the preceding discussion
that in order for the CPV effects in neutrino oscillations 
to be observable, both $\sin(\Delta m^2_{31}L/(2E))$ and
$\sin(\Delta m^2_{21}L/(2E))$ should be sufficiently large.
In the case of  $\sin(\Delta m^2_{31}L/(2E))$, for instance, 
this requires that, say, $\Delta m^2_{31}L/(2E)\sim 1$.
The future experiments on CP violation in 
neutrino oscillations are planned to be performed 
with accelerator $\nu_{\mu}$ and $\bar{\nu}_{\mu}$ beams 
with energies of a few GeV. Taking as an instructive example 
$E = 1$ GeV and using the best fit value of 
$\Delta m^2_{31} = 2.48\times 10^{-3}~{\rm eV^2}$, 
it is easy to check that $\Delta m^2_{31}L/(2E)\sim 1$
for $L \sim 10^{3}$ km. Thus, the study of neutrino 
oscillations requires experiments to have
relatively long baselines. The MINOS, T2K and OPERA experiments 
(see, e.g., Ref. \refcite{PDG2012} 
and references quoted therein), which have provided and 
continue to provide data on $\nu_{\mu}$ oscillations, 
have baselines of approximately 735 km, 295 km and 730 km, 
respectively. The NO$\nu$A experiment, which is under 
preparation and is planned to 
start taking data in 2014, has a baseline of 810 km.

Thus, in the MINOS, OPERA, NO$\nu$A and in the 
future planned experiments 
(see, e.g., Ref.~\refcite{LBLFuture13}) 
the baselines are such that the neutrinos travel 
relatively long distances in the matter of the 
Earth mantle. As is well known, the presence of matter 
can modify drastically the pattern of neutrino 
oscillations\cite{MSW}~. 
When neutrinos propagate in matter, they interact 
with the background of electrons, protons and neutrons, 
which generates an effective potential $V_{eff}$ 
in the neutrino Hamiltonian: $H = H_{vac} + V_{eff}$.
This modifies the neutrino mixing since the eigenstates 
and the eigenvalues  of  $H_{vac}$ and of 
$H = H_{vac} + V_{eff}$ are different, leading to 
different oscillation probabilities with respect 
to those of oscillations in vacuum. 
Typically, the matter background is not
charge conjugation (C-) symmetric: the Earth and the Sun, 
for instance, contain only electrons, protons and neutrons, 
but do not contain their antiparticles.  
As a consequence, the oscillations 
taking place in the Earth, 
are neither CP- nor CPT- invariant \cite{Lang87}~. 
This complicates the studies of
CP violation due to the Dirac phase $\delta$ 
in long baseline neutrino oscillation 
experiments since neutrinos have relatively 
long paths in the Earth 
(see, e.g., Refs.~\refcite{Future,LBLFuture13}).
The matter effects in neutrino oscillations 
in the Earth to a good precision 
are not T-violating\cite{PKSP3nu88} 
since the Earth matter density distribution 
is to a good approximation spherically symmetric.
In matter with constant density, 
e.g., the Earth mantle, one has\cite{PKSP3nu88}~: 
$A^{(e,\mu)}_{\rm T} = J^{\rm m}_{\rm CP}F^{\rm m}_{\rm osc}$,
$J^{\rm m}_{\rm CP} = J_{\rm CP}~R_{\rm CP}$, where 
the dimensionless function $R_{\rm CP}$ does not
depend on $\theta_{23}$ and $\delta$ and 
$|R_{\rm CP}|\ltap 2.5$.  
 
 The expression for the probability
of the $\nu_\mu \rightarrow \nu_{e}$ oscillations
taking place in the Earth mantle in the case
of 3-neutrino mixing, in which both
neutrino mass squared differences
$\Delta m^2_{21}$ and $\Delta m^2_{31}$  contribute
and the CP violation effects due to the Dirac phase
in the neutrino mixing matrix are taken into account,
has the following form in the constant
density approximation and keeping terms
up to second order in the two small
parameters $|\alpha|\equiv |\Delta m^2_{21}|/|\Delta m^2_{31}|\ll 1$
and $\sin^2\theta_{13} \ll 1$~\cite{MFreund04}~:
\begin{equation}
\label{Earthemu}
P^{3\nu~man}_{m}(\nu_{\mu} \rightarrow \nu_{e}) \cong
P_0 + P_{\sin\delta} + P_{\cos\delta} + P_3\,.
\end{equation}
%
\noindent Here
\begin{eqnarray}
\label{P0}
P_0 &= \sin^2\theta_{23}\,\,\frac{\sin^22\theta_{13}}{(A - 1)^2}\,
\sin^2[(A-1)\Delta]\,,
\end{eqnarray}
\begin{eqnarray}
\label{P3}
P_{3} &= \alpha^2\, \cos^2\theta_{23}\,
\frac{\sin^22\theta_{12}}{A^2}\,\sin^2(A\Delta)\,,\\[0.30cm]
\label{Psind}
P_{\sin\delta} &= -\,\alpha \,\, \frac{8\,J_{CP}}{A(1-A)}\,
(\sin\Delta)\,(\sin A\Delta)\, \left (\sin [(1-A)\Delta] \right)\,,\\[0.30cm]
\label{Pcosd}
P_{\cos\delta} &= \alpha \,\, \frac{8\,J_{CP}\, \cot\delta}{A(1-A)}\,
(\cos\Delta)\,(\sin A\Delta)\, \left (\sin [(1-A)\Delta] \right)\,,
\end{eqnarray}
%
\noindent where
\begin{eqnarray}
\alpha = \frac{\Delta m^2_{21}}{\Delta m^2_{31}}\,,
~\Delta = \frac{\Delta m^2_{31}\,L}{4E}\,,~
A = \sqrt{2}G_{\rm F}N^{man}_{e}\frac{2E}{\Delta m^2_{31}}\,,
\label{aDA}
\end{eqnarray}
%
$N^{man}_{e}$ being the electron number density of the Earth mantle.
Thus, the quantity $A$ accounts for the Earth matter effects 
in neutrino oscillations. The mean electron number density
in the Earth mantle is\cite{PREM81}~
$\bar{N}_{e}^{man}\cong 2.2~{\rm cm^{-3}~N_A}$,
${\rm N_A}$ being Avogadro's number. 
In the case of the experiments under discussion, 
the electron number density
$N_e$ changes relatively little around
the indicated mean value along the trajectories of neutrinos
in the Earth mantle and the constant density 
approximation $N_{e}^{man} = const.= \tilde{N}_{e}^{man}$,
$\tilde{N}_{e}^{man}$ 
being the mean density along the
given neutrino path in the Earth,
was shown to be sufficiently accurate in what
concerns the calculation of neutrino oscillation
probabilities\cite{PKSP3nu88,SP3198,SPNu98}~.
The \footnote{The conditions of validity of the analytic 
expression for $P^{3\nu~man}_{m}(\nu_{\mu} \rightarrow \nu_{e})$
given above are discussed in detail in Ref.~\refcite{MFreund04}.}
expression for the $\bar{\nu}_{\mu} \rightarrow \bar{\nu}_{e}$
oscillation probability can be obtained
formally from that for $P^{3\nu~man}_{m}(\nu_{\mu} \rightarrow \nu_{e})$
by making the changes $A\rightarrow -A$ and $J_{CP}\rightarrow - J_{CP}$,
with $J_{CP}\cot\delta \equiv
{\rm Re}(U_{\mu 3}U^*_{e3}U_{e2}U^*_{\mu 2})$
remaining unchanged. The term $P_{\sin\delta}$ in
$P^{3\nu~man}_{m}(\nu_{\mu} \rightarrow \nu_{e})$
would be equal to zero if the Dirac phase
in the neutrino mixing matrix $U$
possesses a CP-conserving value.
Even in this case, however,
we have $A^{(e \mu)~man}_{CP}\equiv
(P^{3\nu~man}_{m}(\nu_{\mu} \rightarrow \nu_{e}) -
P^{3\nu~man}_{m}(\bar{\nu}_{\mu} \rightarrow \bar{\nu}_{e})) \neq 0$
due to the effects of the Earth matter.
It will be important to experimentally
disentangle the effects of the Earth matter
and of $J_{CP}$
in $A^{(e \mu)~man}_{CP}$: this
will allow to get direct information about
the Dirac CP violation phase in $U$.
This can be done, in principle, by studying 
the energy dependence of 
$P^{3\nu~man}_{m}(\nu_{\mu} \rightarrow \nu_{e})$
and  $P^{3\nu~man}_{m}(\bar{\nu}_{\mu} \rightarrow \bar{\nu}_{e})$.
In the vacuum limit of $N^{man}_e = 0$ ($A = 0$)
we have $A^{(e \mu)~man}_{CP} = A^{(e \mu)}_{CP}$
(see Eq. (\ref{CPTAsym})) and only the term 
$P_{\sin\delta}$ contributes
to the asymmetry $A^{(e \mu)}_{CP}$. 
 
 The preceding remarks apply also to the probabilities  
$P^{3\nu~man}_{m}(\nu_{e} \rightarrow \nu_{\mu})$  and
$P^{3\nu~man}_{m}(\bar{\nu}_{e} \rightarrow \bar{\nu}_{\mu}))$.
The probability $P^{3\nu~man}_{m}(\nu_{e} \rightarrow \nu_{\mu})$, 
for example, can formally be obtained from the expression 
for the probability $P^{3\nu~man}_{m}(\nu_{\mu} \rightarrow \nu_{e})$ 
by changing the sign of the term  $P_{\sin\delta}$.

%
\subsection{Majorana CP Violation Phases and $\betabeta$-Decay}
%

The massive neutrinos $\nu_j$ can be Majorana 
fermions. Many theories of neutrino mass generation 
predict massive neutrinos to be Majorana fermions 
(see, e.g., Refs.~\refcite{seesaw,ThRMoh05,STP82PD}). 
If $\nu_j$ are proven to be 
Majorana particles, the neutrino mixing matrix $U$, 
as we have already emphasised,  
will contain two additional 
CP violation ``Majorana'' phases\cite{BHP80}~,
$\alpha_{21}$ and $\alpha_{31}$.
Getting experimental information about the  
Majorana CPV phases $\alpha_{21}$ and $\alpha_{31}$
in $U$ will be remarkably difficult
\cite{BPP1,PPW,BargerCP,PPR1,PPSchw05,MajPhase1}~.
The oscillations of flavour neutrinos, 
$\nu_{l} \rightarrow \nu_{l'}$ and 
$\bar{\nu}_{l} \rightarrow \bar{\nu}_{l'}$,
$l,l'=e,\mu,\tau$, are insensitive to the phases
$\alpha_{21,31}$~\cite{BHP80,Lang87}~.
The phases $\alpha_{21,31}$ 
can affect significantly the predictions for 
the rates of the (LFV) decays $\mu \rightarrow e + \gamma$,
$\tau \rightarrow \mu + \gamma$, etc.
in a large class of supersymmetric theories
incorporating the see-saw mechanism\cite{PPY03}~. 
As we will discuss further, 
the Majorana phase(s) in the PMNS matrix can
play the role of the leptogenesis 
CPV parameter(s) at the origin of the baryon 
asymmetry of the Universe\cite{PPRio106}~. 

 The Majorana nature of massive neutrinos manifests 
itself in the existence of processes in which the 
total lepton charge changes by two units, 
$|\Delta L|= 2$: 
$K^+ \rightarrow \pi^- + \mu^+ + \mu^+$,
$e^- +(A,Z) \rightarrow e^+ + (A,Z-2)$, etc.
The only feasible experiments which
at present have the potential of
establishing the Majorana nature 
of light neutrinos $\nu_j$ and of
providing information on the Majorana 
CPV phases in PMNS matrix are the experiments 
searching for neutrinoless double beta
($\betabeta$-) decay, 
$(A,Z) \rightarrow (A,Z+2) + e^- + e^-$,
of even-even nuclei $^{48}$Ca, $^{76}$Ge, 
$^{82}$Se, $^{100}$Mo, $^{116}$Cd, $^{130}$Te, 
$^{136}$Xe, $^{150}Nd$, etc. (see, 
e.g., Refs.~\refcite{STPNuNature2013,Hindawibb0nu2013}). 
In  $\betabeta$-decay, two neutrons of the initial nucleus 
$(A,Z)$ transform by exchanging the virtual light massive Majorana 
neutrino(s) $\nu_j$ into two protons of the final state 
nucleus $(A,Z+2)$ and two free electrons. 
The corresponding $\betabeta$-decay amplitude has the 
form (see, e.g., Refs.~\refcite{BiPet87,WRodej10}):
$A(\betabeta) = G^2_{\rm F}\, \mefff\, M(A,Z)$, 
where  $G_{\rm F}$ is the Fermi constant, $\mefff$ is 
the  $\betabeta$-decay effective Majorana mass and 
$M(A,Z)$ is the nuclear matrix element (NME) of the process.
The $\betabeta$-decay effective Majorana mass $\mefff$  
contains all the dependence of the 
$\betabeta$-decay amplitude on the neutrino 
mixing parameters. 
We have (see, e.g., Refs.~\refcite{BiPet87,WRodej10}):
\begin{equation}
\meff=\left| m_1 |U_{\mathrm{e} 1}|^2 
+ m_2 |U_{\mathrm{e} 2}|^2e^{i\alpha_{21}}
 + m_3 |U_{\mathrm{e} 3}|^2e^{i(\alpha_{31} - 2\delta)} \right|\,,
\label{effmass2}
\end{equation}
%
\noindent $|U_{\mathrm{e}1}|$=$c_{12}c_{13}$,
$|U_{\mathrm{e}2}|$=$s_{12}c_{13}$, 
$|U_{\mathrm{e}3}|$=$s_{13}$. 
For the normal hierarchical (NH), inverted hierarchical (IH) 
and quasi-degenerate (QD) neutrino mass spectra 
$\meff$ is given by (see, e.g., Ref.~\refcite{STPFocusNu04}): \\ 
\begin{figure}[htb]
\vspace{-0.5cm}
 \includegraphics[width=10.0cm,height=8.0cm,clip=]{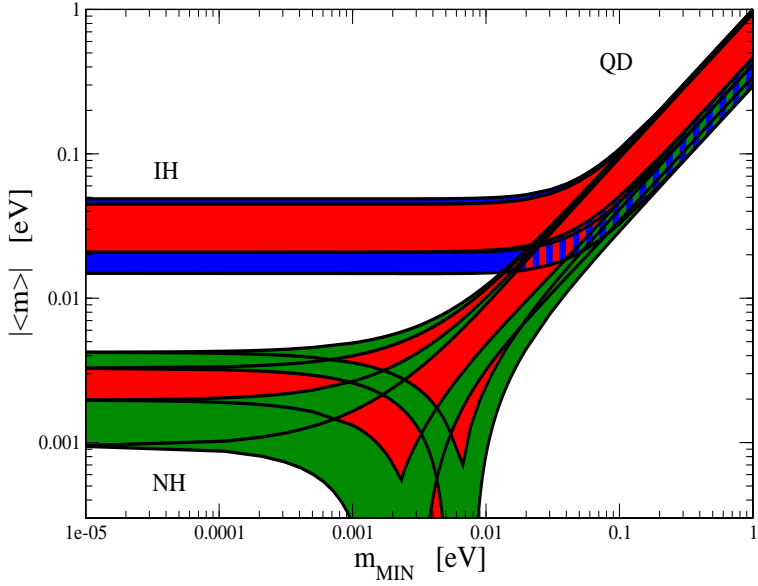}
\vskip -0.4cm
\caption{
The effective Majorana mass $\meff$
(including a 2$\sigma$ uncertainty),
as a function of $m_{\rm min}= {\rm min}(m_j)$ 
for $\sin^2\theta_{13} = 0.0236 \pm 0.0042$~\protect\cite{DBayth13}, 
$\delta =0$ and using the 95\% C.L. allowed ranges of 
$\Delta m^2_{21}$, $|\Delta m^2_{31(32)}|$,  
$\sin^2 \theta_{12}$ and $\sin^2 \theta_{13}$ 
found in Ref.~\refcite{fogli11}.
The phases $\alpha_{21,31}$
are varied in the interval [0,$\pi$].
The predictions for the NH, IH and QD
spectra are indicated.
The red regions correspond to at least one of
the phases $\alpha_{21,31}$
and $(\alpha_{31} - \alpha_{21})$
having a CP violating value, while the 
blue and green areas correspond to $\alpha_{21,31}$ 
possessing CP conserving values. 
(From Ref.~\protect\refcite{PDG2012}.)
}
\label{Fig1}
\vspace{-0.3cm}
\end{figure}
%

\noindent 
$\meff\cong |\sqrt{\Delta m^2_{21}}~s^2_{12}
+\sqrt{\Delta m^2_{31}}~s^2_{13}e^{i(\alpha_{32}-2\delta)}|$,~NH,\\ 

\noindent
$\meff \cong \sqrt{|\Delta m^2_{32}|}~
\left|c^2_{12} + s^2_{12}~e^{i\alpha_{21}}\right|$,~~~IH,\\

\noindent
$\meff \cong m_0~\left|c^2_{12}
 + s^2_{12}~e^{i \alpha_{21}}\right|$,~~~QD,\\

\noindent where $\alpha_{32}$=$\alpha_{31}$-$\alpha_{21}$.
Obviously, $\meff$ depends strongly 
on the Majorana phase(s):
the CP-conserving values of 
$\alpha_{21}$=$0,\pm\pi$ \cite{LW81}~, 
for instance, determine the range of 
possible values of $\meff$ in the 
cases of IH and QD spectrum.
As is well-known, if CP-invariance holds,
the phase factor\\ 

\vspace{-0.3cm}
$\eta_{jk} = e^{i\alpha_{jk}} = \pm 1$,~$j>k$,~$j,k=1,2,3$,\\
%

\vspace{-0.3cm}
\noindent represents \cite{LW81,BiPet87}
the relative CP-parity of 
Majorana neutrinos $\nu_{j}$ and $\nu_k$,\\
\noindent
$\eta_{jk} = \eta^{\nu CP}_{j}(\eta^{\nu CP}_k)^{\ast}$,
$\eta^{\nu CP}_{j(k)} = \pm i$ being the CP-parity of
$\nu_{j(k)}$.

Using the $3\sigma$ ranges of the allowed values of 
the neutrino oscillation parameters 
quoted in Eqs. (\ref{theta13values}) - (\ref{deltasolvalues}),
one finds that:\\
i) $0.70\times 10^{-3}~{\rm eV} \ltap \meff
\ltap 4.51\times 10^{-3}$ eV 
in the case of NH spectrum;\\
ii) $1.4\times 10^{-2}~{\rm eV} \ltap \meff 
\ltap 4.8\times 10^{-2}$ eV in the case of IH spectrum;\\
iii) $2.8\times 10^{-2}~{\rm eV} \ltap \meff \ltap m_0$ eV,
$m_0\gtap 0.10$ eV, in the case of QD spectrum.\\ 
The difference in the ranges of 
$\meff $ in the cases of NH, IH and QD spectrum opens
up the possibility to get information about
the type of neutrino mass spectrum
from a measurement of 
$\meff$~\cite{PPSNO2bb}~.
The  main features of the 
predictions for $\meff$   
are illustrated in Fig. 1, 
where $\meff$ is shown as a function of the lightest 
neutrino mass $m_{\rm min} \equiv {\rm min}(m_j)$.

  The experimental searches for $\betabeta$-decay  
have a long history (see, e.g., Ref.~ \refcite{bb0nuExp1}).
A positive $\betabeta$-decay signal at $> 3\sigma$,
corresponding to $T_{1/2}^{0\nu} = (0.69 - 4.18)\times
10^{25}$~yr (99.73\%~C.L.) and implying $\meff = (0.1 -
0.9)~{\rm eV}$, is claimed to have been observed in 
\cite{KlapdorMPLA}~, while a later analysis\cite{Klap04} reports 
evidence for $\betabeta$-decay at 6$\sigma$ 
with $T_{1/2}^{0\nu}(^{76}Ge) =  2.23^{+0.44}_{-0.31}\times 10^{25}~\text{yr}$,
corresponding to $\meff = 0.32 \pm 0.03$~eV. 
The best lower limit on the half-life of $^{76}$Ge,
$T_{1/2}^{0\nu}(^{76}Ge) > 2.1\times 10^{25}$~yr (90\%~C.L.),
was found in the GERDA $^{76}$Ge 
experiment\cite{GERDAGe762013}~. 
By combining the limits obtained in the 
Heidelberg-Moscow\cite{HMGe76}~, 
IGEX\cite{IGEX00}~ and GERDA  
experiments one gets\cite{GERDAGe762013}~ 
$T_{1/2}^{0\nu}(^{76}Ge) > 3.0\times 10^{25}$~yr (90\%~C.L.).

  Two experiments, NEMO3~\cite{NEMO3}
with $^{100}$Mo and CUORICINO~\cite{CUORI}  
with $^{130}$Te,
obtained the limits: 
$\meff < (0.61 - 1.26)$ eV~~\cite{NEMO3} 
and $\meff< (0.16 - 0.68)$ eV~~
\cite{CUORI} (90\% C.L.),
where estimated uncertainties in the NME 
are accounted for.
The best lower limits on the $\betabeta$-decay half-life 
of $^{136}$Xe were reported by the EXO~\cite{EXO2012}
and KamLAND-Zen~\cite{KLZen2012} collaborations: 
$T_{1/2}^{0\nu}(^{136}Xe) > 1.6\times 10^{25}$~yr~~\cite{EXO2012} 
and $T_{1/2}^{0\nu}(^{136}Xe) > 1.9\times 10^{25}$~yr~~\cite{KLZen2012} 
(90\%~C.L.).

 Most importantly, a large number of 
experiments of a new generation aim at
sensitivity to $\meff \sim (0.01 - 0.05)$ eV (see, e.g., 
Ref.~\refcite{Hindawibb0nu2013}): CUORE
($^{130}$Te), GERDA ($^{76}$Ge), SuperNEMO, EXO ($^{136}$Xe), MAJORANA
($^{76}$Ge), AMoRE ($^{100}$Mo), MOON ($^{100}$Mo), COBRA ($^{116}$Cd), 
CANDLES ($^{48}$Ca), KamLAND-Zen ($^{136}$Xe), 
SNO+ ($^{130}Te$), etc.  GERDA, EXO 
and  KamLAND-Zen have provided already 
the best lower limits on the $\betabeta$-decay
half-lives of $^{76}$Ge and $^{136}$Xe. 
The experiments listed above
are aiming to probe the QD and IH ranges of $\meff$; 
they will test the positive result claimed 
in Ref.~\refcite{Klap04}. 
If the $\betabeta$-decay will be 
observed in these experiments, 
the measurement of the $\betabeta$-decay 
half-life might allow to obtain constraints on 
the Majorana phase $\alpha_{21}$~ 
\cite{BGKP96,BPP1,PPW}~ (see also Ref.~\refcite{fogli11}). 
 
The possibility of establishing  CP
violation in the lepton sector 
due to Majorana CPV  phases has been studied 
in Refs.~\refcite{PPW,BargerCP} and in 
much greater detail in Refs.~\refcite{PPR1,PPSchw05}~.
It was found that it is very challenging:
it requires quite accurate measurements 
of $\meff$ (and of $m_0$ for QD spectrum), 
and holds only for a limited range of 
values of the relevant parameters.
More specifically~\cite{PPR1,PPSchw05}~, 
establishing at 2$\sigma$
CP-violation associated with
Majorana neutrinos 
in the case of QD spectrum requires 
for $\sin^2\theta_{12}$=0.31, in particular, 
a relative experimental 
error on the measured value of 
$\meff$ and $m_0$ smaller than 15\%,
a ``theoretical uncertainty'' 
$F{\small \ltap}$1.5 in the value of
$\meff$ due to an imprecise 
knowledge of the corresponding NME, 
and value of the relevant Majorana
CPV phase $\alpha_{21}$ typically within the ranges 
of ${\small\sim (\pi/4 - 3\pi/4)}$ and
${\small\sim (5\pi/4 - 7\pi/4)}$. 

The knowledge of NME with 
sufficiently small uncertainty 
\footnote{A possible test of 
the NME calculations is suggested in Ref.~\refcite{PPW} and 
is discussed in greater detail in Ref.~\refcite{NMEBiPet04}~ 
(see also, e.g., Ref. \refcite{NME2012})} 
is crucial for obtaining quantitative information on
the neutrino mixing parameters from a measurement of
$\betabeta$-decay half-life 
\footnote{For discussions of the current status of the 
calculations of the NMEs for the $\betabeta$-decay see, e.g., 
the third article quoted in Ref.~\refcite{Hindawibb0nu2013} and 
Ref.~\refcite{NME2012}.}. 
The observation of
a $\betabeta$-decay of one nucleus is likely to lead
to the searches and eventually to observation
of the decay of other nuclei.
One can expect that such a progress, in particular,
will help to solve completely the problem
of the sufficiently precise calculation
of the nuclear matrix elements
for the $\betabeta$-decay~\cite{PPW}~.

 If the future $\betabeta$-decay experiments 
show that $\meff < 0.01$ eV, both the IH and
the QD spectrum will be ruled out for massive 
Majorana neutrinos. If in addition it is 
established in neutrino oscillation 
experiments that the neutrino mass spectrum is 
with {\it inverted ordering}, i.e. 
that $\Delta m^2_{31(32)} < 0$,
one would be led to conclude that 
either the massive neutrinos $\nu_j$ 
are Dirac fermions, or that 
$\nu_j$ are Majorana particles
but there are additional contributions to the 
\betabeta-decay amplitude which 
interfere destructively 
with that due to the exchange of 
light massive Majorana neutrinos. 
The case of more than one mechanism 
generating the $\betabeta$-decay was discussed 
recently in, e.g., 
Refs.~\refcite{AMMultiple11}~,
where the possibility to identify the mechanisms 
inducing the decay was also analised.
If, however, $\Delta m^2_{31(32)}$ 
is determined to be positive
in neutrino oscillation experiments, 
the upper limit $\meff < 0.01$ eV 
would be perfectly compatible with 
massive Majorana neutrinos
possessing NH mass spectrum, 
or mass spectrum with normal ordering but
partial hierarchy, and the quest for $\meff$ 
would still be open.

 Let us emphasise that determining the 
nature of massive 
neutrinos is one of the fundamental, 
most challenging and pressing problems 
in today's neutrino physics
(see, e.g. Refs.~\refcite{PDG2012,Hindawibb0nu2013}). 
Establishing whether the neutrinos with definite mass
$\nu_j$ are Dirac fermions possessing distinct antiparticles, 
or Majorana fermions, i.e., spin 1/2 particles that 
are identical with their antiparticles, is 
of fundamental importance for understanding 
the origin of neutrino masses and mixing and 
the underlying symmetries of particle 
interactions (see, e.g., Ref.~\refcite{ThRMoh05}).
We recall that the neutrinos $\nu_j$ 
will be Dirac fermions if the 
particle interactions conserve 
some additive lepton number, e.g., the total
lepton charge $L = L_e + L_{\mu} + L_{\tau}$. 
If no lepton charge is conserved, 
the neutrinos $\nu_j$ will be Majorana fermions. 
As we have seen, the massive neutrinos 
$\nu_j$ are predicted to be of Majorana nature
by the see-saw mechanism 
\cite{seesaw}~. 
The observed patterns of neutrino mixing 
and of neutrino mass squared differences 
can be related to Majorana massive neutrinos 
and the existence of an 
{\it approximate} flavour symmetry in the 
lepton sector (see, e.g., Ref.~\refcite{STP82PD}).
Determining the nature (Dirac or Majorana)
of massive neutrinos $\nu_j$
is one of the major goals of 
the program of research in 
neutrino physics.

\vspace{-0.2cm}
%
\section{The See-Saw Mechanism and Leptogenesis}
%

  A natural explanation of the smallness of
neutrino masses is provided by the see-saw mechanism 
of neutrino mass generation\cite{seesaw}~. 
An integral part of the simplest version of this mechanism - 
the so-called ``type I see-saw'', are the $SU(2)_L$ singlet 
RH neutrinos $\nu_{lR}$, $l=e,\mu,\tau$. 
The latter are assumed to possess 
a Majorana mass term as well as Yukawa type coupling 
with the Standard Theory lepton and Higgs doublets
$\psi_{lL}(x)$ and $\Phi(x)$, respectively,
$(\psi_{lL}(x))^T = (\nu^T_{lL}(x)~~l^T_{L}(x))$, $l=e,\mu,\tau$,
$(\Phi(x))^T = (\Phi^{(0)}~\Phi^{(-)})$.
The Standard Theory admits such a minimal extension 
which does not modify any of the basic attractive 
features of the Theory (unitarity, renormalisability, etc.).
In the basis in which the Majorana mass 
matrix of RH neutrinos is diagonal we have: 
\beq
\label{Ynu}
{\cal L}_{\rm Y,M}(x) =  
 - ( \lambda_{kl}\,\overline{N_{kR}}(x)\, \Phi^{\dagger}(x)\,\psi_{lL}(x)
 + \hbox{h.c.} )  
 -\,\frac{1}{2}\,M_{k}\,\overline{N_k}(x)\, N_k(x)\,,
\eeq
%
where $\lambda_{lk}$ is the matrix of neutrino 
Yukawa couplings and $N_k(x)$ is the heavy (RH) 
Majorana neutrino field possessing a mass $M_k > 0$, 
$M_1 < M_2 < M_3$. The fields $N_k(x)$ satisfy 
the Majorana condition 
$C \overline{N_k}^T(x) = \xi_k  N_k(x)$,
where $C$ is the charge conjugation matrix 
and $\xi_k$ is a phase.
When the electroweak 
symmetry is broken spontaneously, 
the neutral component 
of the Higgs doublet field develops non-zero 
vacuum expectation value 
$v = 174$ GeV and 
the neutrino Yukawa coupling generates 
a neutrino Dirac mass term: 
$m^{D}_{kl}\,\overline{N_{kR}}(x)\, \nu_{lL}(x) + \hbox{h.c.}$,
with $m^{D} = v\lambda$. 
In the case when the elements of 
$m^{D}$ are much smaller than $M_k$,
$|m^{D}_{jl}|\ll M_k$, $j,k=1,2,3$, $l=e,\mu,\tau$, 
the interplay between the
Dirac mass term and the Majorana mass term 
of the heavy singlets $N_k$ generates an 
effective Majorana mass 
(term) for the LH flavour neutrino fields   
$\nu_{lL}(x)$~~\cite{seesaw}~:
\begin{equation}
(m^{\nu})_{l'l}\cong v^2(\lambda^T\,M^{-1}\,\lambda)_{l'l} 
= ( (m^{D})^T\,M^{-1}\,m^{D} )_{l'l}
= (U^{\ast}\, m \,U^{\dagger})_{l'l}\,,
\label{mnu}
\end{equation}
%
where $M \equiv {\rm Diag}(M_1,M_2,M_3)$ 
($M_{1,2,3} > 0$),
$m \equiv {\rm Diag}(m_1,m_2,m_3)$,
$m_j \geq 0$ being the mass of the 
light Majorana neutrino $\nu_j$,
and $U$ is the PMNS matrix
The diagonalisation of the
mass matrix $m^{\nu}$
leads to the appearance of the PMNS 
neutrino mixing matrix in the 
charged current
weak interaction Lagrangian 
${\cal L}_{\rm CC}(x)$, Eq. (\ref{CC}).

 In grand unified theories, $m^{D}$ is typically of the
order of the charged fermion masses. 
In $SO(10)$ theories~\cite{seesaw}~, 
for instance, $m^{D}$ coincides with the up-quark mass matrix.
Taking indicatively $m^{\nu} \sim$ 0.05 eV, 
$m^{D}\sim$ 100 GeV, one finds 
$M_k\sim 2\times 10^{14}$ GeV, which is close 
to the scale of unification of electroweak and
strong interactions, $M_{GUT}\cong 2\times 10^{16}$ GeV.   
In GUT theories with RH neutrinos one finds 
that indeed the heavy  singlets $N_k$ naturally 
obtain masses which are  
by few to several orders of magnitude smaller 
than $M_{GUT}$ (see, e.g., Ref.~\refcite{ThRMoh05}).

 One of the characteristic predictions of the 
see-saw mechanism is that both the light and 
heavy neutrinos $\nu_j$ and $N_k$ are Majorana particles. 
As we have discussed, the Majorana nature of the 
light neutrinos can be revealed in the $\betabeta$- 
decay experiments.

  We will discuss next briefly the 
interesting possibility~\cite{PPRio106,EMSTP09} 
that the CP violation necessary 
for the generation of the baryon 
asymmetry of the Universe, $Y_B$,
in the leptogenesis scenario
can be due exclusively to 
the Dirac and/or Majorana CPV phases 
in the PMNS matrix, and thus can be 
directly related to the 
low energy leptonic CP violation 
(e.g., in neutrino oscillations, etc.). 
We recall that leptogenesis~\cite{LGFY} is 
a simple mechanism which allows 
to explain the observed 
baryon asymmetry of the Universe~\cite{PlanckYB2013}~, 
namely the observed difference in the present epoch of the 
evolution of the Universe of the 
number densities of baryons and anti-baryons,
$n_{\rm B}$ and $n_{\rm \bar{B}}$ : 
\begin{equation}
Y_{\rm B} = \frac{n_{B} - n_{\bar{B}}}{s_0} = 
(8.67 \pm 0.15)\times 10^{-11}\,, 
\label{YB0}
\end{equation}
%
where $s_0$ is the entropy density 
in the current epoch 
\footnote{The entropy density $s$ at temperature $T$ 
is given by $s = g_{*}(2\pi^2/45)T^3$, 
where $g_{*}$ is the number of (thermalised) 
degree of freedom at temperature $T$. In the 
present epoch of the evolution of the Universe 
we have $s_0 = 7.04\,n_{\gamma 0}$,
$n_{\gamma 0}$ being the number density of photons.
}. 
The simplest scheme in which the leptogenesis 
mechanism can be implemented is the 
type I see-saw model. 
In its minimal version it includes the 
Standard Theory plus two or three   
heavy (RH) Majorana neutrinos, $N_k$.
Thermal leptogenesis (see, e.g., Ref.~\refcite{lept}) 
can take place, e.g., in the case of 
hierarchical spectrum of the heavy neutrino masses, 
$M_1 \ll M_2 \ll M_3$, which we consider 
in what follows. The lepton asymmetry is 
produced in the Early Universe in out-of-equilibrium lepton 
number and CP nonconserving decays of the lightest heavy  
Majorana neutrino, $N_1$,
mediated by the neutrino Yukawa couplings, 
$\lambda$. The lepton asymmetry is  
converted into a baryon asymmetry
by $(B-L)$-conserving but
$(B+L)$-violating sphaleron interactions~ 
\cite{kuzmin}
which exist within the Standard Theory and are efficient at 
temperatures $T\gtap 100$ GeV. 
In grand unified theories the 
heavy neutrino masses fall  typically in the range of 
$\sim(10^8 - 10^{14})$ GeV 
(see, e.g., Ref.~\refcite{ThRMoh05}). 
This range coincides with the range of 
values of $M_k$, required for a successful 
thermal leptogenesis~\cite{lept}~.
For hierarchical spectrum of the heavy neutrino masses
$M_1 \ll M_2 \ll M_3$ we consider, 
leptogenesis takes place in the Early Universe 
typically at temperatures somewhat smaller than the 
mass of $N_1$, but not smaller than roughly $10^9$ GeV, 
$10^9~{\rm GeV}\ltap T < M_1$.

  In our further discussion it is 
convenient to use the ``orthogonal 
parametrisation`` of the matrix 
of neutrino Yukawa couplings~\cite{Casas:2001sr}~:
\begin{eqnarray}
\label{R}
\lambda = v^{-1} \, \sqrt{M} \, R\, \sqrt{m}\, U^{\dagger},~~
R~R^T = R^T~R = {\bf 1},
\end{eqnarray} 
%
\noindent where $R$ is, in general, a complex matrix.
It is parametrised, in general, by six real parameters 
(e.g., three complex angles), of which three 
parameters can have CP violating values. 
 
In the setting we are considering the only source of CP 
violation in the lepton sector is the matrix of neutrino 
Yukawa couplings $\lambda$. It is clear from Eq. (\ref{R}) 
that the CP violating parameters in the matrix $\lambda$ 
can have their origin from the CP violating  
phases in the PMNS matrix $U$, or from the  CP violating 
parameters present in the matrix $R$, or else 
from both the CP violating parameters in $U$ and in $R$.

 For determining the conditions under which
the CP-violation responsible for 
leptogenesis is due exclusively to 
the Dirac and/or Majorana CPV 
phases in the PMNS matrix, 
it is useful to analyze the constraints 
which the requirement of  CP-invariance 
imposes on the Yukawa couplings 
$\lambda_{j l}$, on the PMNS matrix 
$U$ and on the matrix $R$. 
These constraints read (in a certain well specified 
and rather widely used convention)~~\cite{PPRio106}~:
\begin{equation}
\lambda^{\ast}_{j l}=\lambda_{j l}\,\rho^{N}_j\,,~~\rho^{N}_j = \pm 1\,,
~~j=1,2,3,~l=e,\mu,\tau\,,
\label{YCPinv2}
\end{equation}
\begin{equation}
U^{\ast}_{l j} = U_{l j}\,\rho^{\nu}_j\,,~~\rho^{\nu}_j = \pm 1\,,~~
j=1,2,3,~l=e,\mu,\tau\,,
\label{UCPinv1}
\end{equation}
\begin{equation}
 R^{\ast}_{jk} = R_{jk}\,\rho^{N}_j \,\rho^{\nu}_k\,,~~j,k=1,2,3\,,
\label{RCPinv1}
\end{equation}
\noindent where $i\,\rho^{N}_j = \pm i$ and $i\,\rho^{\nu}_k =\pm i$ 
are the CP-parities of the heavy and light 
Majorana  neutrinos $N_j$ and $\nu_k$ 
(see, e.g., Refs. \refcite{STPNuNature2013,BiPet87}). 
Obviously, the last would be a condition
of reality of the matrix $R$ only if 
$\rho^{N}_j \rho^{\nu}_k$=1 for any $j,k$=1,2,3.
However, we can also have 
$\rho^{N}_j \rho^{\nu}_k$=$- 1$
for some $j$ and $k$ and in that case 
$R_{jk}$ will be {\it purely imaginary}. 
Of interest for our further analysis is, 
in particular, the product
\begin{equation}
P_{jkml}\equiv R_{j k}\,R_{j m} \, U^{\ast}_{lk} \, U_{lm}\,,~ k\neq m\,.
\label{P1}
\end{equation}
%
\noindent If CP-invariance holds, we find 
from the conditions given above 
that $P_{jkml}$ has to be real~~\cite{PPRio106}~:
 \begin{equation}
P^{\ast}_{jkml} = P_{jkml}\,(\rho^{N}_j)^2\,
(\rho^{\nu}_k)^2\, (\rho^{\nu}_m)^2 = P_{jkml}\,.
\label{P2}
\end{equation}
%
\noindent Consider the case
when CP-invariance conditions 
for the PMNS matrix are satisfied and 
$U^{\ast}_{\tau k}U_{\tau m}$ for 
given $k$ and $m$, $k<m$, $k=1,2$, $m=2,3$, 
is purely imaginary, i.e.,
${\rm Re}(U^{\ast}_{\tau k} U_{\tau m})$=0.
This can be realised
for $\delta = \pi q$, $q$=0,1,2, and 
$\rho^{\nu}_k \rho^{\nu}_m = -1$, 
i.e., if the relative CP-parity of the light Majorana
neutrinos $\nu_k$ and  $\nu_m$ is equal to $(-1)$,
or, correspondingly, if 
$\alpha_{mk}$=$\pi (2q' +1)$, $q'$=0,1,....
In this case CP-invariance 
holds in the lepton sector at ``low'' energies. 
In order for CP-invariance to hold 
at ``high'' energy, i.e., 
for $P_{jkml}$ to be real,
the product $R_{j k} R_{j m}$
has also to be {\it purely imaginary},
${\rm Re}(R_{j k} R_{j m}) = 0$.
Thus, in the case considered, 
{\it purely imaginary} $U^{\ast}_{\tau k}U_{\tau m}\neq 0$
and {\it real} $R_{j k} R_{j m} \neq 0$, 
i.e., ${\rm Re}(U^{\ast}_{\tau k} U_{\tau m})=0$,
${\rm Im}(R_{j k} R_{j m}) = 0$,
in particular, 
{\it imply violation of CP-symmetry at 
``high'' energy  by the interplay of the matrices $U$ and $R$}.
 
  The realization that the CP violation necessary 
for the generation of the baryon 
asymmetry of the Universe can be due exclusively to the
CPV phases in the PMNS matrix, is related to the 
progress in the understanding of the 
importance of lepton flavour effects in leptogenesis
~\cite{davidsonetal,nardietal}
(for earlier discussion see Ref.~\refcite{Barbieri99}).
In the case of hierarchical heavy neutrinos $N_k$,
$M_1 \ll M_2 \ll M_3$, the flavour effects in leptogenesis 
can be significant for~\cite{davidsonetal,nardietal} 
$10^{8}~{\rm GeV} \ltap  M_1 \ltap (0.5 - 1.0)\times 10^{12}$ GeV.
If the requisite lepton asymmetry is produced in 
this regime, the CP violation necessary for 
successful leptogenesis can be provided entirely by the 
CPV phases in the neutrino mixing matrix~~\cite{PPRio106}~. 

  Indeed, suppose that the 
mass of $N_1$ lies in the interval of interest,
$10^{9}~{\rm GeV} \ltap M_1 \ltap 10^{12}$ GeV.
The CP violation necessary for the generation of 

\begin{figure}
  \centerline{
 {\includegraphics[width=10truecm,height=8cm]{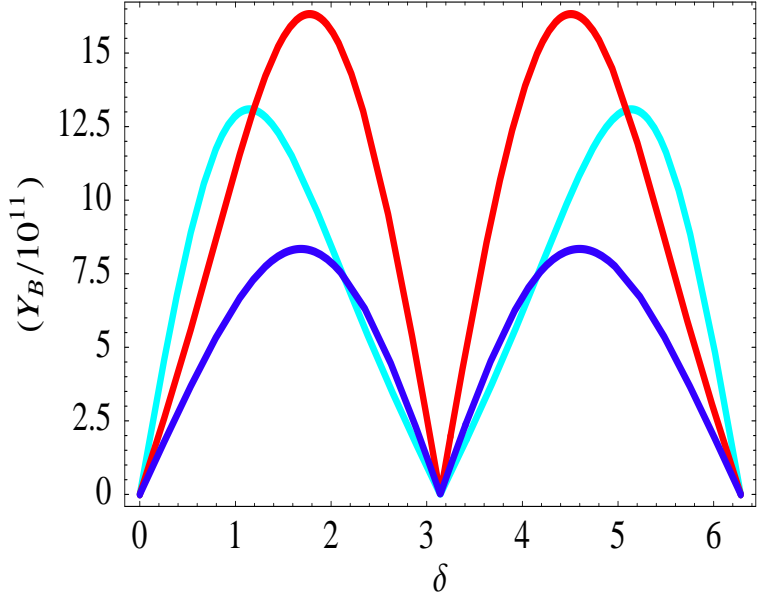}}}
  \caption{ 
The baryon asymmetry $|Y_B|$ as a 
function of the Dirac phase $\delta$ 
varying in the interval $\delta = [0,2\pi]$ in the case of 
Dirac CP-violation, $\alpha_{32}=0;~2\pi $, 
hierarchical heavy neutrinos and 
NH light neutrino mass spectrum, 
for $M_1 = 5\times 10^{11}$ GeV,
real $R_{12}$ and $R_{13}$ satisfying
$|R_{12}|^2 + |R_{13}|^2 = 1$,
$|R_{12}| =0.86$, $|R_{13}| =0.51$, 
${\rm sign}\left(R_{12}R_{13}\right)=+1$, and
for i) $\alpha_{32}=0$,
$s_{13}=0.2$ (red line) and    
$s_{13}=0.1$ (dark blue line), 
ii) $\alpha_{32}= 2\pi$, 
$s_{13}=0.2$ (light blue line).  
(From Ref.~\refcite{PPRio106}.)
}
 \label{Figs678241106.eps}
 \end{figure}
%

\noindent the baryon asymmetry $Y_B$ in ``flavoured'' leptogenesis 
can arise, as we have already noted, both 
from the ``low energy'' neutrino mixing 
matrix $U$ and/or from the 
 ``high energy'' part of the matrix of 
neutrino Yukawa couplings $\lambda$ - 
the matrix $R$, which can mediate CP violating 
phenomena  only at some high energy scale 
determined by the masses $M_k$ of the heavy Majorana 
neutrinos $N_k$. The matrix $R$ 
does not affect the ``low'' 
energy neutrino mixing phenomenology.
Suppose further that the matrix $R$
has real and/or purely imaginary 
CP-conserving elements: we are interested in the 
case when the CP violation necessary for 
leptogenesis is due exclusively to the 
CPV phases in  $U$. Under these assumptions, 
$Y_B$ generated via leptogenesis can be written as~ 
\cite{davidsonetal,nardietal}
\begin{equation}
|Y_B|\cong 3\times 10^{-3}~|\epsilon_{\tau}~\eta|\,, 
\label{YB}
\end{equation}
%
where $\epsilon_{\tau}$ is the 
CPV asymmetry in the $\tau$ flavour 
(lepton charge)
produced in $N_1$-decays
\footnote{We have given the expression
for $Y_B$ normalised to the entropy density,
see, e.g., Ref.~\refcite{PPRio106}.}, 
\begin{eqnarray}
\epsilon_{\tau} = -\,\frac{3 M_1}{16\pi v^2}\frac{{\rm Im}(
\sum_{j k} 
m_j^{1/2}m_k^{3/2} U^*_{\tau j}U_{\tau k}
R_{1j}R_{1k})}{\sum_i m_i |R_{1i}|^2}\,,
\label{epsa1}
\end{eqnarray}
%
\begin{figure}[htb]
\vskip -0.5cm
\includegraphics[width=10cm,height=7cm,clip=]{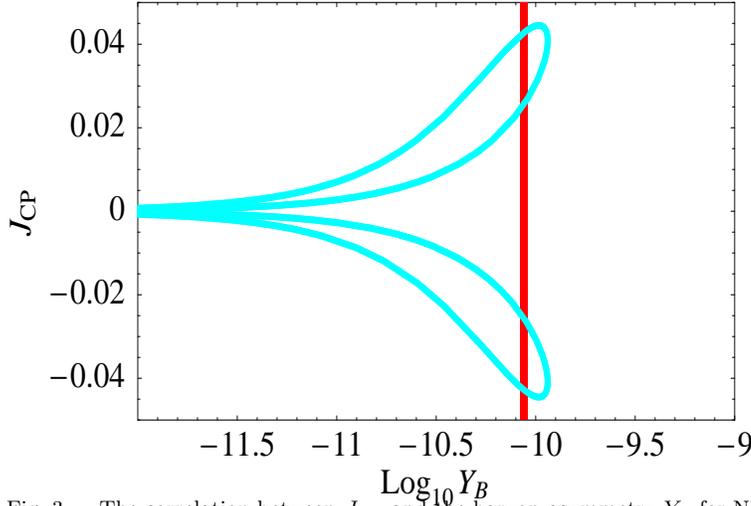}
\vskip -0.6cm
\caption{
The correlation between $J_{\rm CP}$   
and the baryon asymmetry $Y_{\rm B}$
for NH spectrum,
$s_{13}$=0.2, $\alpha_{32}$=0, 
$R_{12}$=0.86 
and $M_1$=$5\times 10^{11}$ GeV.
The Dirac phase is varied in the interval
$\delta = [0,2\pi]$.
The $2\sigma$ allowed range of $Y_{\rm B}$
is also shown. (From Ref.~\refcite{PPRio106}.)
}
\label{Fig1}
\end{figure}
%

\noindent $\eta$ is the efficiency factor~\cite{davidsonetal}~,
\begin{equation}
|\eta| \cong |\eta(0.71\widetilde{m_2}) - 
\eta(0.66\widetilde{m_\tau})|\,,
\label{eta}
\end{equation}
%
$\widetilde{m_{2,\tau}}$ 
being the wash-out mass parameters 
which determine the rate of the processes in the Early 
Universe that tend to ``erase'', or ``wash-out'', 
the asymmetry,
\begin{equation}
\widetilde{m_2}= \widetilde{m_e} + \widetilde{m_\mu}\,,~~
\widetilde{m_l} = |\sum_{j} m_j~R_{1j}~U^*_{l j}|^2\,,~~l=e,\mu\,.
\label{woutms}
\end{equation}
%
\noindent Approximate analytic expression for 
$\eta(\widetilde{m})$ is given in~ 
\cite{davidsonetal,nardietal}~.
We shall consider next a few specific examples.\\

\vspace{-0.3cm}
{\it A. NH Spectrum, $m_1 \ll m_2 \ll m_3\cong \sqrt{\Delta m^2_{31}}$}.\\

\indent Assume for simplicity that $m_1\cong 0$ and
$R_{11}\cong 0$ ($N_3$ decoupling). If $R_{12}R_{13}$ 
is real and $\alpha_{32}$=0, the only source of 
CP-violation is the Dirac phase $\delta$ 
 in $U$, and $\epsilon_{\tau}
\propto \sin\theta_{13}\sin\delta$. 
For $R_{12}R_{13} > 0$,
$s_{13}=0.15$, $\delta$=$3\pi/2$,  
and $R_{12}\cong 0.86$ 
(which maximises $|Y_B|$), we have 
~\cite{PPRio106}~:
$|Y_B| \cong 2.7\times 10^{-13}$
($\sqrt{\Delta m^2_{31}}/0.05~{\rm eV}) 
(M_1/10^9~{\rm GeV})$, 
where we have used the 
best fit values of $\Delta m^2_{21}$, 
$\sin^2\theta_{12}$ and 
$\sin^2\theta_{23}$ 
(see Fig. \ref{Figs678241106.eps}). 
For the values of 
$M_1\ltap 5\times 10^{11}~{\rm GeV}$
for which the flavour effects in leptogenesis 
can be significant, the observed 
value of the baryon asymmetry,  
taken conservatively to lie in the interval
$|Y_B| \cong (8.1 - 9.3)\times 10^{-11}$,
can be reproduced if 
\begin{equation}
|\sin\theta_{13}\sin\delta|\gtap 0.09\,,~~{\rm or}~ 
|J_{\rm CP}|\gtap 2.0\times 10^{-2}\,.
\label{BAUNHth13delta}
\end{equation}
%
The ranges of values of 
$\sin\theta_{13}$ and of $|J_{\rm CP}|$
we find  in the case being studied
are comfortably compatible with the measured value 
of $\sin\theta_{13}$ and with 
the hints that $\delta \cong 3\pi/2$.  
Since both $Y_B$ and $J_{\rm CP}$
depend on $s_{13}$ and $\delta$,
for given values of the other 

\begin{figure}[htb]
\vskip -0.5cm
\includegraphics[width=10cm,height=7cm,clip=]{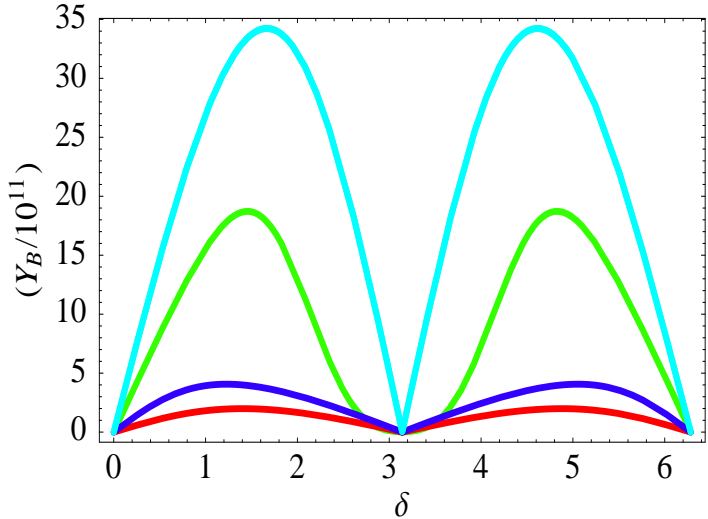}
\vskip -0.30cm
 \caption{ 
The asymmetry $|Y_B|$ as a function 
of the Dirac phase $\delta$ 
in the case of hierarchical heavy  neutrinos,  
IH light neutrino mass spectrum, 
Dirac CP-violation,  
$\alpha_{21} = \pi$ 
$R_{11}R_{12} = i~\kappa~|R_{11}R_{12}|$ 
($|R_{11}|^2 - |R_{12}|^2 = 1$),
$\kappa = - 1$ (red and dark blue lines),
$\kappa = + 1$ (light blue and green lines), 
for $M_1 = 2\times 10^{11}$ GeV,
and  $s_{13} = 0.1$ (red and green lines)
and $s_{13} = 0.2$ (dark blue and light blue lines). 
Values of $|R_{11}|$, 
which maximise $|Y_B|$ 
have been used: $|R_{11}| = 1.05$ 
in the case of $\kappa = - 1$, 
and $|R_{11}| = 1.3~(1.6)$  
for $\kappa = + 1$ and $s_{13} = 0.2~(0.1)$. 
 (From Ref.~\refcite{PPRio106}.)
}
\label{Fig13IHDalphapi040107Delt}
\vspace{-0.3cm}
\end{figure}
%

\noindent relevant parameters there exists 
a correlation between the values of  
$|Y_B|$ and $J_{\rm CP}$. This correlation 
is illustrated in Fig. \ref{Fig1}.

As was shown in~\cite{PPRio106}~,
we can have successful leptogenesis 
also if the sole source of CP-violation 
is the difference of the Majorana phases 
$\alpha_{32}$=$\alpha_{31}-\alpha_{21}$ 
of $\pmns$. In this case values of 
$M_1\gtap 4\times 10^{10}~{\rm GeV}$ 
are required.\\

\vspace{-0.3cm}
{\it B. IH Spectrum, $m_3 \ll 
m_{1,2}\cong \sqrt{|\Delta m^2_{32}|}$.}\\

\begin{figure}[htb]
\vskip -0.5cm
\includegraphics[width=10.0cm,height=6.8cm,clip=]{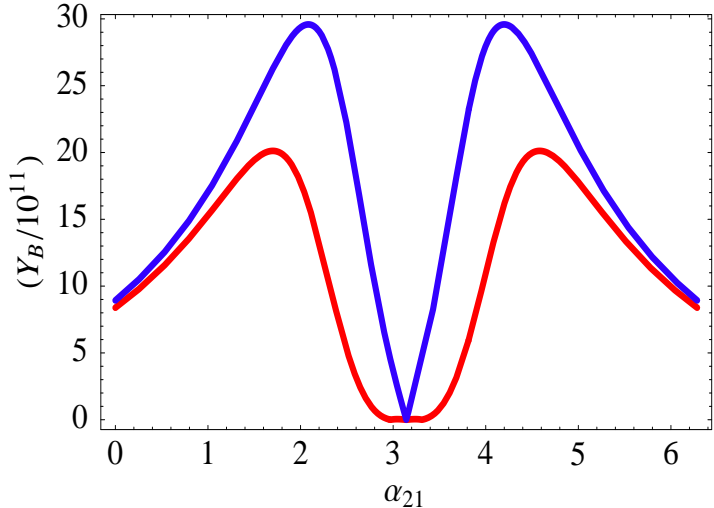}
\vskip -0.30cm
\caption{
The asymmetry $|Y_B|$ versus 
the Majorana phase $\alpha_{21} = [0,2\pi]$, 
for IH spectrum, purely imaginary
$R_{11}R_{12} = i\kappa |R_{11}R_{12}|$, 
$\kappa = - 1$,
$|R_{11}|^2 - |R_{12}|^2 = 1$,
$M_1 = 2\times 10^{11}$ GeV,
$\delta = 0$ and
$s_{13}=0~(0.2)$ - blue (red) line. 
 (From Ref.~\refcite{PPRio106}.)
}
\label{IHs130bl02ralpha21minus.eps}
\vspace{-0.3cm}
\end{figure}

\indent Under the simplifying 
conditions of $m_3 \cong 0$ 
and $R_{13}\cong 0$ ($N_3$ decoupling), 
leptogenesis can be successful for 
$M_1\ltap 10^{12}~{\rm GeV}$
only if $R_{11}R_{12}$ is not real~\cite{PRST05,PPRio106}, 
so we consider the case of  purely imaginary 
$R_{11}R_{12}$=$i\kappa |R_{11}R_{12}|$, 
$\kappa$=$\pm 1$. The requisite 
CP-violation can be due to the
i) Dirac phase $\delta$ (Fig. \ref{Fig13IHDalphapi040107Delt}), 
and/or ii) Majorana phase 
$\alpha_{21}$ (Fig. \ref{IHs130bl02ralpha21minus.eps}), 
in the neutrino mixing matrix $U$. If, 
e.g., in the second case 
we set $\sin\delta=0$ (say, $\delta = \pi$), the maximum of 
$|Y_B|$ for, e.g., $\kappa$=$-1$, 
is reached for~\cite{PPRio106}~ 
$|R_{11}|^2\cong 1.4$ 
($|R_{12}|^2 = |R_{11}|^2 - 1=0.4$),
and $\alpha_{21}\cong 2\pi/3;4\pi/3$, 
and at the maximum 
$|Y_B| \cong 1.5\times 10^{-12}
(\sqrt{|\Delta m^2_{32}|}/(0.05~{\rm eV}) 
(M_1/10^9~{\rm GeV})$.
The observed $|Y_B|$ can be reproduced for 
$M_1 \gtap 5.4\times 10^{10}~{\rm GeV}$.
Since both $|Y_B|$  and the effective 
Majorana mass in $\betabeta$-decay,
$\meff$, depend on the Majorana phase 
$\alpha_{21}$, there exists a correlation 
between the values of $|Y_B|$ and $\meff$.
%
\noindent

 Similar results can be obtained~\cite{PPRio106}
in the case of quasi-degenerate in mass
heavy Majorana neutrinos.

  The interplay in ``flavoured'' leptogenesis
between contributions in $Y_B$ due 
to the ``low energy'' and ``high energy'' 
CP violation, originating from the 
PMNS matrix $U$ and the $R$-matrix, respectively, 
was investigated in Ref.~\refcite{EMSTP09}~.
It was found, in particular, that 
under certain conditions which can 
be tested in low energy neutrino 
experiments (IH spectrum, 
$(-\sin\theta_{13}\cos \delta)\gtap 0.1$),
the ``high energy'' contribution in $Y_B$ 
due to the $R$-matrix, can be so 
strongly suppressed that it would play 
practically no role in the generation 
of baryon asymmetry compatible with the observations. 
One would have successful leptogenesis 
in this case only if the requisite CP violation 
is provided by the Majorana phases 
in the PMNS matrix $U$. 

\vspace{-0.2cm}
%
\section{Conclusions}
%
%

 The program of research in neutrino physics aims at shedding light 
on some of the fundamental aspects of neutrino mixing:\\

i) the nature of massive neutrinos 
$\nu_j$, which can be Dirac fermions possessing 
distinct antiparticles, or Majorana fermions, 
i.e., spin 1/2 particles that are identical with 
their antiparticles;\\

ii) the type of spectrum the neutrino masses obey;\\

iii) the status of CP symmetry in the lepton sector;\\

iv) the absolute scale of neutrino masses.\\

The program extends beyond the year 2025 
(see, e.g., Refs.~\refcite{SnowM2013,LBLFuture13}).
Our ultimate goal is to understand at a fundamental level
the mechanism giving rise to neutrino masses and mixing and to
non-conservation of the lepton charges $L_l$, $l=e,\mu,\tau$. 
This includes understanding the origin of the
patterns of neutrino mixing and of neutrino masses
suggested by the data. The remarkable experimental 
program of research in neutrino physics 
(the cost of which is expected to exceed 
altogether 1.3 billion US dollars)
and the related theoretical efforts are stimulated 
by the fact that the existence of nonzero neutrino masses and 
the smallness of the neutrino masses suggest the existence of new 
fundamental mass scale in particle physics, 
i.e., the existence of New Physics beyond 
that predicted by the Standard Theory.
It is hoped that progress in the theory of
neutrino mixing will also lead, in particular, 
to progress in the theory of flavour and  
to a better understanding of the
mechanism of generation of the baryon
asymmetry of the Universe. 
%

\vspace{-0.2cm}
\section*{Acknowledgements.}
This work was supported in part by the Italian 
INFN program on ``Fisica Astroparticellare''. Partial 
support from the Organising Committee of the Conference 
is acknowledged with gratefulness.


\vspace{-0.4cm}
\begin{thebibliography}{9}
%
\bibitem{PDG2012} 
K. Nakamura and S.~T. Petcov,
in J. Beringer {\it et al.} (Particle Data Group),
{\em Phys. Rev. D} {\bf 86}, 010001 (2012).
%
\bibitem{STPNuNature2013} S.T. Petcov, 
{\em  Adv. High Energy Phys.}  {\bf 2013}, 852987 (2013)
  [arXiv:1303.5819].
%
\bibitem{BPont67}
  B.~Pontecorvo,
  {\em Zh. Eksp. Teor. Fiz.}  {\bf 53}, 1717 (1967).
%

\bibitem{LGFY}
 M.~Fukugita and T.~Yanagida,
 {\em Phys. Lett. B} {\bf 174}, 45 (1986).
%
\bibitem{kuzmin}
V.A. Kuzmin {\it et al.},
{\em Phys. Lett. B} {\bf 155}, 36 (1985).
%
\bibitem{seesaw}
P.~Minkowski,
 {\em  Phys. Lett. B} {\bf 67}, 421 (1977); 
M. Gell-Mann, P. Ramond and R. Slansky in Sanibel Talk,
CALT-68-709, Feb 1979, and in {\it Supergravity} (North Holland,
Amsterdam 1979);
T. Yanagida in {\it Proc. of the Workshop on Unified Theory and
Baryon Number of the Universe}, KEK, Japan, 1979;
S.L.Glashow, Cargese Lectures (1979).
%
\bibitem{BHP80} S.M. Bilenky, J. Hosek and S.T. Petcov,
             {\em  Phys. Lett. B} {\bf 94}, 495 (1980).
%
\bibitem{PPRio106} S. Pascoli {\it et al.}, 
{\em Nucl. Phys. B} {\bf 774}, 1 (2007);
see also {\em Phys. Rev. D} {\bf 75}, 083511 (2007).
%
\bibitem{EMSTP09} E. Molinaro, S.T. Petcov,
{\em Phys. Lett. B} {\bf 671}, 60  (2009); arXiv:0803.4120. 
%
\bibitem{BPont57} B. Pontecorvo, 
                  {\em Zh. Eksp. Teor. Fiz.} 
{\bf 33}, 549 (1957) and {\bf 34}, 247 (1958). 
%
\bibitem{MNS62} Z. Maki, M. Nakagawa and S. Sakata, 
{\em Prog. Theor. Phys.} {\bf 28}, 870 (1962).
%

\bibitem{SterNuWhitePaper} K.N. Abazajian {\it et al.}, 
arXiv:1204.5379; see also T. Lasserre, talk given at 
TAUP2013, September 9-13, 2013, Asilomar, California, U.S.A. 
%
\bibitem{SnowM2013} A. de Gouvea {\it et al.}, arXiv:1310.4340.
%
\bibitem{BiPet87} S.~M.~Bilenky and S.~T.~Petcov,
  {\em Rev. Mod. Phys.}  {\bf 59}, 671 (1987).
%
\bibitem{Capozzi:2013csa}
  F.~Capozzi {\it et al.},
  arXiv:1312.2878.
%
\bibitem{CGGMSchw12update} 
M. C. Gonzalez-Garcia {\it et al.},
{\em JHEP} {\bf 12}, 123 (2012);
the updated results obtained after the TAUP2013 International 
Conference (held in September of 2013) are posted at the URL 
www.nu-fit.org/?q=node/45.
%
\bibitem{DBayth13}  F.P. An {\it et al.},
{\em Phys. Rev. Lett.} {\bf 108}, 171803 (2012); 
{\em Chinese Phys.} {\bf C37}, 011001 (2013); arXiv:1310.6732.
%
\bibitem{RENOth13}  J.K. Ahn {\it et al.}, 
{\em Phys. Rev. Lett.} {\bf 108}, 191802 (2012); 
 S.-H. Seo [for the RENO Collab.], talk at the
TAUP2013 International Workshop, 
September 9-13, 2013, Asilomar, California, U.S.A.
%
\bibitem{PPNH07} S. Pascoli and S.T. Petcov  
{\em Phys. Rev. D} {\bf 77}, 113003 (2008).
%
\bibitem{NOIO}
~S.T. Petcov, M. Piai,  
{\em Phys. Lett. B} {\bf 533}, 94  (2002).
S. Pascoli and S.T. Petcov, 
{\em Phys. Lett. B} {\bf 544}, 239 (2002); 
J. Bernab\'eu {\it et al.}, 
{\em Nucl. Phys. B} {\bf 669}, 255 (2003);
~S.T. Petcov and S. Palomares-Ruiz,
{\em Nucl. Phys. B} {\bf 712}, 392 (2005);
S.T. Petcov, T. Schwetz, 
{\em Nucl. Phys. B} {\bf 740}, 1 (2006).
%
\bibitem{Cahn2013}
R.N. Cahn  {\it et al.}, arXiv:1307.5487.
%
\bibitem{Fermi34} 
F. Perrin, {\em Comptes Rendus} {\bf 197}, 868 (1933);
E. Fermi,  {\em Nuovo Cim.} {\bf 11}, 1 (1934).
%
\bibitem{Mainz} Ch. Kraus {\it et al.},
{\em Eur. Phys. J.} {\bf C40}, 447 (2005). 
%
\bibitem{MoscowH3} V. Lobashev {\it et al.},   
{\em Nucl. Phys. A} {\bf 719}, 153c  (2003).
%
\bibitem{MoscowH3b} V.N. Aseev {\it et al.},
{\em Phys. Rev.} D {\bf 84}, 112003 (2011).
%
\bibitem{MainzKATRIN} K. Eitel {\it et al.},
{\em Nucl. Phys. B} (Proc. Suppl.) {\bf 143}, 197 (2005). 
%
\bibitem{summj} K.N. Abazajian {\it et al.},
{\em Astopart. Phys.} {\bf 35}, 177 (2011). 
%
\bibitem{Ade:2013lta} P.A.R. Ade {\it et al.}, [Planck Collab.], 
arXiv:1303.5076, to be published in {\em Astrophys. J}.
%
\bibitem{WMAPascitedbyPlanck} C. L. Bennett {\it et al.}, 
arXiv:1212.5225, to be published in {\em Astrophys.\ J.\ Supp.} 
%
\bibitem{ACTascitedbyPlanck} J. Dunkley {\it et al.},
{\em JCAP} {\bf 1307}, 025 (2013).
%
\bibitem{Marzocca:2013cr} D.~Marzocca {\it et al.},
{\em JHEP} {\bf 1305}, 073 (2013) 
[arXiv:1302.0423].  
%
\bibitem{Cabibbo78} N. Cabibbo, 
{\em Phys. Lett. B} {\bf 72}, 333 (1978).
%
\bibitem{PKSP3nu88}
P.I. Krastev and S.T. Petcov, 
{\em Phys. Lett. B} {\bf 205}, 84 (1988).
%
\bibitem{Barger:1980jm}
  V.~D.~Barger, K.~Whisnant and R.~J.~N.~Phillips,
  Phys.\ Rev.\ Lett.\  {\bf 45}, 2084 (1980). 
%
\bibitem{CJ85} C. Jarlskog, {\em Z. Phys. C} {\bf 29}, 491 (1985).
%
\bibitem{LBLFuture13}
S.K. Agarwalla {\it et al.}, arXiv:1312.6520;
C. Adams {\it et al.}, arXiv:1307.5700.
%
\bibitem{MSW} L.~Wolfenstein, 
{\em Phys. Rev. D} {\bf 17}, 2369 (1978), and
{\em Proc. of the 8th International Conference on Neutrino Physics 
and Astrophysics - ``Neutrino'78"},
(ed. E.C. Fowler, Purdue University Press, 
West Lafayette, 1978), p. C3; 
S.P. Mikheev and A.Y. Smirnov,
{\em Soviet J. Nucl. Phys.} {\bf 42}, 913 (1885), and
Nuovo Cimento {\bf 9C}, 17 (1986); 
see also: V. Barger {\it et al.}, 
{\em Phys. Rev. D} {\bf 22}, 2718 (1980).
%
\bibitem{Lang87} P. Langacker {\it et al.},  
{\em Nucl. Phys. B} {\bf 282}, 589 (1987).

\bibitem{Future} A. Bandyopadhyay {\it et al.}, 
{\em Rept. Prog. Phys.} {\bf 72}, 106201 (2009).
%
\bibitem{MFreund04} M. Freund, 
{\em Phys. Rev. D} {\bf 64}, 053003 (2001). 
%
\bibitem{PREM81} A.D.~Dziewonski and D.L.~Anderson, 
{\em Physics of the Earth and Planetary Interiors} 
               {\bf 25}, 297 (1981).
%
\bibitem{SP3198}  S.T.~Petcov, 
{\em Phys. Lett. B} {\bf 434}, 321 (1998), 
(E) {\it ibid.} {\bf B444}, 584 (1998);
see also: Nucl.\ Phys.\ (Proc. Suppl.) {\bf B77}, 93 (1999) and 
hep-ph/9811205.
%
\bibitem{SPNu98} M.V.~Chizhov, M.~Maris, and S.T.~Petcov, 
hep-ph/9810501.
%
\bibitem{ThRMoh05} R. Mohapatra {\it et al.},
{\em Rept. Prog. Phys.} {\bf 70}, 1757 (2007);
A. Bandyopadhyay {\it et al.}, 
{\em Rept. Prog. Phys.} {\bf 72}, 106201 (2009);
S. King and Ch. Luhn, arXiv:1301.1340,
and references quoted therein.
%
\bibitem{STP82PD} 
S.T. Petcov, {\em Phys. Lett. B} {\bf 110}, 245 (1982); 
~P.H. Frampton, S.T. Petcov and W. Rodejohann, 
{\em Nucl. Phys. B} {\bf 687}, 31 (2004);
 I. Girardi  {\it et al.}, 
JHEP {\bf 1402}, 050 (2014),
and references quoted therein.
%
\bibitem{BPP1} S.M. Bilenky, S. Pascoli and S.T. Petcov,
              {\em Phys.\ Rev.} D{\bf 64}, 053010 (2001). 
%
\bibitem{PPW}  S. Pascoli, S.T. Petcov and L. Wolfenstein,
            {\em Phys. Lett. B} {\bf 524}, 319 (2002).
%
\bibitem{BargerCP} V.~Barger  {\it et al.}, 
{\em Phys. Lett. B} {\bf 540}, 247 (2002). 
%
\bibitem{PPR1} S. Pascoli, S.T. Petcov and W. Rodejohann 
{\em Phys. Lett. B}  {\bf 549}, 177 (2002).
%
\bibitem{PPSchw05} S. Pascoli, S.T. Petcov and T. Schwetz, 
{\em Nucl. Phys. B} {\bf 734}, 24 (2006).
%
\bibitem{MajPhase1} A. De Gouvea, B. Kayser, R. Mohapatra,
{\em Phys. Rev. D} {\bf 67}, 053004 (2003). 

\bibitem{PPY03} S. Pascoli {\it et al.}, 
{\em Phys. Lett. B} {\bf 564}, 241 (2003).
%
\bibitem{Hindawibb0nu2013} C. Aalseth {\it et al.}, hep-ph/0412300;
F. Avignone, {\em Nucl. Phys. Proc. Suppl.} {\bf 143}, 233 (2005); 
A. Giuliani and A. Poves, {\em Advances in High Energy Physics} 
{\bf 2012}, 857016 (2012). 
%
\bibitem{WRodej10} 
W. Rodejohann, 
{\em Int. J. Mod. Phys. E} {\bf 20}, 1833 (2011).
%
\bibitem{STPFocusNu04}
 S.T. Petcov, 
{\em Physica Scripta} {\bf T121}, 94 (2005). 
%
\bibitem{LW81} L. Wolfenstein, 
{\em Phys. Lett. B} {\bf 107}, 77 (1981); 
~S.M. Bilenky {\it et al.}, 
{\em Nucl. Phys. B} {\bf 247}, 61 (1984); 
B. Kayser, {\em Phys. Rev. D} {\bf 30}, 1023 (1984).
%
\bibitem{PPSNO2bb} S. Pascoli, S.T. Petcov,
{\em Phys. Lett. B} {\bf 544}, 239 (2002).
%
\bibitem{bb0nuExp1} A. S. Barabash, 
{\em  Phys. Atom. Nucl.} {\bf 74}, 603 (2011). 

\bibitem{KlapdorMPLA}
  H.~V.~Klapdor-Kleingrothaus {\it et al.},
{\em  Mod.\ Phys.\ Lett.},  A16:2409, 2001.

\bibitem{Klap04} H.V. Klapdor-Kleingrothaus  
{\it et al.},
{\em Phys. Lett. B} {\bf 586}, 198 (2004)
%
\bibitem{GERDAGe762013} K.-H. Ackermann {\it et al.}, 
 {\em Phys. Rev. Lett.} {\bf 111}, 122503 (2013).
%
\bibitem{HMGe76}
H.V. Klapdor-Kleingrothaus {\it et al.}, 
{\em Nucl. Phys. Proc. Suppl.} {\bf 100}, 309 (2001). 
%
\bibitem{IGEX00} C.E. Aalseth {\it et al.},
{\em Phys. Atomic Nuclei} {\bf 63}, 1225 (2000). 
%
\bibitem{NEMO3} A. Barabash {\it et al.},
{\em J. Phys. Conf. Ser.} {\bf 173}, 012008 (2009).
%
\bibitem{CUORI} C. Amaboldi {\it et al.},
{\em Phys. Rev. C} {\bf 78}, 035502 (2008).
%
\bibitem{EXO2012} M. Auger {\it et al.},
{\em Phys. Rev. Lett.} {\bf 109}, 032505 (2012).

\bibitem{KLZen2012}  
  A.~Gando {\it et al.},
  {\em Phys.  Rev.  Lett.}  {\bf 110}, 062502 (2013). 


\bibitem{BGKP96} S.M. Bilenky {\it et al.}, 
{\em Phys. Rev. D} {\bf 56}, 4432 (1996).

\bibitem{fogli11} G.L. Fogli {\it et al.},
{\em Phys. Rev. D} {\bf 84}, 053007 (2011).
%
\bibitem{NMEBiPet04} S.M. Bilenky and S.T. Petcov, 
hep-ph/0405237.

\bibitem{NME2012} J. Vegados, H. Ejiri and F. Simkovi\^c, 
{\em Rept. Progr. Phys.} {\bf 75}, 106301 (2012). 
%
\bibitem{AMMultiple11} A. Faessler {\it et al.},
{\em Phys. Rev. D} {\bf 83}, 113003 (2011);
A.~Meroni, S.~T.~Petcov and F. Simkovi\^c,
{\em JHEP} {\bf 1302}, 025 (2013).
%
\bibitem{PlanckYB2013} P.A.R. Ade {\it et al.}, [Planck Collab.], 
arXiv:1303.5076, to be published in {\em Astrophys.\ J}.
%
\bibitem{lept}
W. Buchmuller {\it et al.}, {\em Annals Phys.} 
{\bf 315}, 305  (2005). 
%
\bibitem{Casas:2001sr}
J.A. Casas and A. Ibarra,  
{\em Nucl. Phys. B} {\bf  618}, 171 (2001).
%
\bibitem{davidsonetal}
A. Abada {\it et al.}, {\em JCAP} {\bf 0604}, 004 (2006);
{\em JHEP} {\bf 0609}, 010 (2006).
%

\bibitem{nardietal}
E. Nardi {\it et al.}, {\em JHEP} {\bf 0601}, 164 (2006). 

\bibitem{Barbieri99} R. Barbieri {\it et al.}, 
{\em Nucl. Phys. B} {\bf 575}, 61  (2000).
%
\bibitem{PRST05}
S.T.~Petcov {\it et al.}, 
{\em Nucl. Phys. B} {\bf 739}, 208 (2006).

\end{thebibliography}
\end{document}